\documentclass[A4, twocolumn]{article}

\usepackage{graphicx}
\usepackage{dcolumn}
\usepackage{bm}

\usepackage{amsthm}
\usepackage{numprint}
\usepackage{subcaption}
\usepackage{hyperref}
\usepackage{fancyhdr}
\usepackage{csquotes}
\usepackage{chronosys}
\usepackage{amsmath}
\usepackage{amsfonts}
\usepackage{color}
\usepackage{chngcntr}
\usepackage{stmaryrd}
\usepackage{tcolorbox}
\usepackage{floatrow}

\usepackage{authblk}
 
\usepackage{caption}

\usepackage{lineno}

\DeclareCaptionLabelSeparator{dot}{.\space}
\newcommand{\cor}[1]{\textcolor{black}{#1}}

\newcommand{\corsec}[1]{\textcolor{black}{#1}}

\begin{document}

\title{Modeling Newtonian noise of acoustic origin \\ in the Virgo gravitational wave detector}

\author[1]{Lionel Maurin}
\author[1]{François Gautier}
\author[1]{Maël Brun}
\author[1]{Soizic Terrien}

\author[2]{Matteo Barsuglia}

\author[3]{Donatella Fiorucci}

\author[4]{Irene Fiori}
\author[4]{Maria Tringali}

\author[5]{Roberto Passaquieti}

\author[6]{Federico Paoletti}

\author[7]{Mariusz Suchenek}

\author[8]{Tomasz Bulik}

\affil[1]{Laboratoire d'Acoustique de l'Université du Mans (LAUM),
UMR 6613, Institut d'Acoustique - Graduate School (IA-GS), CNRS, Le Mans Université, France}
\affil[2]{Université Paris Cité, CNRS, AstroParticule et Cosmologie, F-75013 Paris, France}
\affil[3]{ENEA, Frascati Research Center, Via E. Fermi, 45, 00044 Frascati, Italy}
\affil[4]{European Gravitational Observatory, EGO, Via E. Amaldi, 56021 Cascina, Pisa, Italy}
\affil[5]{INFN, Sezione di Pisa, I-56127 Pisa, Italy, Università di Pisa, I-56127 Pisa, Italy}
\affil[6]{Istituto Nazionale di Fisica Nucleare, INFN, Sezione di Pisa, Via F. Buonarroti, 3, 56127 Pisa, Italy}
\affil[7]{Nicolaus Copernicus Astronomical Center, Polish Academy of Sciences, Bartycka 18, 00-716, Warsaw, Poland}
\affil[8]{Astronomical Observatory,  University of Warsaw, Aeje ujazdoskie 4, 00478 Warsaw Poland}

\date{August 17, 2026}

\twocolumn[
\begin{@twocolumnfalse}
\maketitle
\begin{abstract}

Since the first gravitational-wave (GW) detection of September 14th 2015 and with hundreds of gravitational-wave sources identified by the LIGO-\cor{Virgo}-KAGRA network, GW have produced many important results in astrophysics and fundamental physics. Along with planned new data takings, current detectors will be upgraded and new project, such as Einstein Telescope and Cosmic Explorer, are under study. Among noises limiting low frequency sensitivity, vibro-acoustic noises are particularly important. In this work, we focus on the gravity gradient noise (also called Newtonian noise) of acoustic origin, which refers to the small fluctuations in the gravity field resulting from the acoustic pressure field present in the experimental areas of the detector. The induced noise is quantified in an original way, using a detailed numerical acoustic model of the experimental room, when the pressure field is excited by the air conditioning system. The method is used for Virgo, but it can be easily extended for future detectors and used to guide the design of caverns and experimental areas. 

\end{abstract}
\end{@twocolumnfalse}
]

\section{Introduction\label{sec:1-intro}}

\begin{figure}
    \centering
    \includegraphics[width = \linewidth]{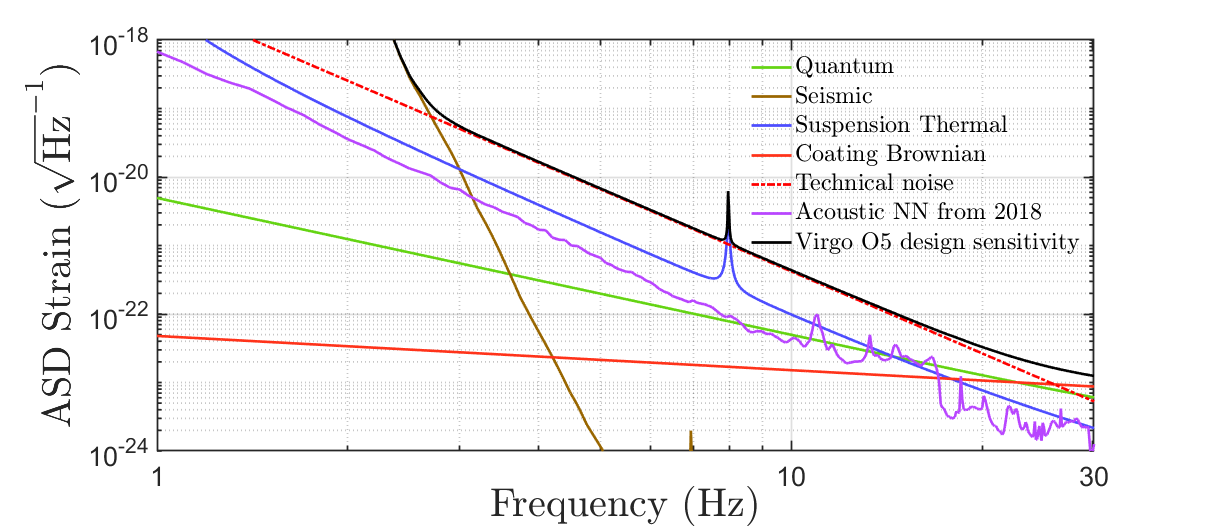}
    \caption{Noise budget generated using the gwinc suite assuming 80 W of input power \cite{gwinc}. The most contributing low frequency noise sources are indicated on the graph. The purple curve is the estimated airborne acoustic Newtonian noise as computed in \cite{Fiorucci_2018} and is the most recent estimate of this noise (2018).}
    \label{fig:1}
\end{figure}

The LIGO, Virgo, and KAGRA interferometric detectors \cite{Virgo_2015,Ligo_2015,Kagra_2020} have released catalogs with hundreds of gravitational-waves (GW) sources \cite{LIGOScientific:2025slb}, producing important results in the fields of astrophysics of compact objects, cosmology, and tests of general relativity. The future scientific potential of GW science is directly related to the capability of improving the detector's sensitivity. Characterizing and reducing the various noises impacting the detectors thus becomes critical. If, in the mid to high part of the spectrum (above 20 $\mathrm{Hz}$), thermal noise and quantum noise are the main contributors to the sensitivity, the low frequency part is limited by a forest of technical and environmental noises. Among them, the acoustic noise produced by variations in the pressure field of the experimental areas (See Fig. \ref{fig:1}). It can impact detectors, either by modulating stray light in non isolated optical components or by producing a change in air density and, therefore, in the gravitational field around the test masses (mirrors) of the detector. This gravitational anomaly translates into a parasitic force that can limit the sensitivity of the interferometer. Such noise, called Newtonian noise (NN) has been extensively modeled when it is produced by a seismic source \cite{Harms_TerrestrialGravityFluctuations_2019,Cafaro_2009, Saulson_1984}. Newtonian noise of acoustic origin has received less attention. Note that this noise cannot be shielded, but mitigation strategies have been proposed. Ideas such as building the detector underground and/or subtracting offline the noise using an array of seismometers around the detector are actively being researched and developed \cite{Maria_2020,Trozzo_2022,Koley_2024}. 

In this study, we focus on the airborne Newtonian noise caused by acoustic modes inside the experimental room of Virgo's end buildings. At low frequencies (1-30 $\mathrm{Hz}$), the sound sources in both end buildings include the Heating, Ventilation and Air Conditioning (HVAC) systems. This complex aero-acoustic source gives rise to acoustic perturbations inside the buildings and must be characterized properly. Determining the importance of the Newtonian noise induced by this specific technical equipment in the end buildings requires the combination of the mathematical framework on Newtonian noise \cite{Harms_TerrestrialGravityFluctuations_2019} with a numerical acoustic model of the room. The paper is divided into two sections. 

In the first section, the North End Building and the West End Building (NEB and WEB respectively) are described. Then, classical room acoustic indicators that are the Schroeder frequency and the acoustic absorption coefficient are derived from previous measurements performed on-site. Lastly, the mathematical framework used for the computation of the gravity field using a functional basis associated to the acoustic modes is described.

In the second section, we present the results of the study: a comparison between the simulated pressure field and the measured pressure field is first used to validate the model. A simplified version of this model is then proposed to reduce computation time. Finally, the induced Newtonian noise at the location of the test mass is calculated and analyzed.

\section{Modeling Newtonian noise \label{sec:2-NNmain} of Acoustic origin}
\subsection{Description of Virgo's terminal building}

\begin{figure}
    \centering
    \includegraphics[width=0.6\linewidth]{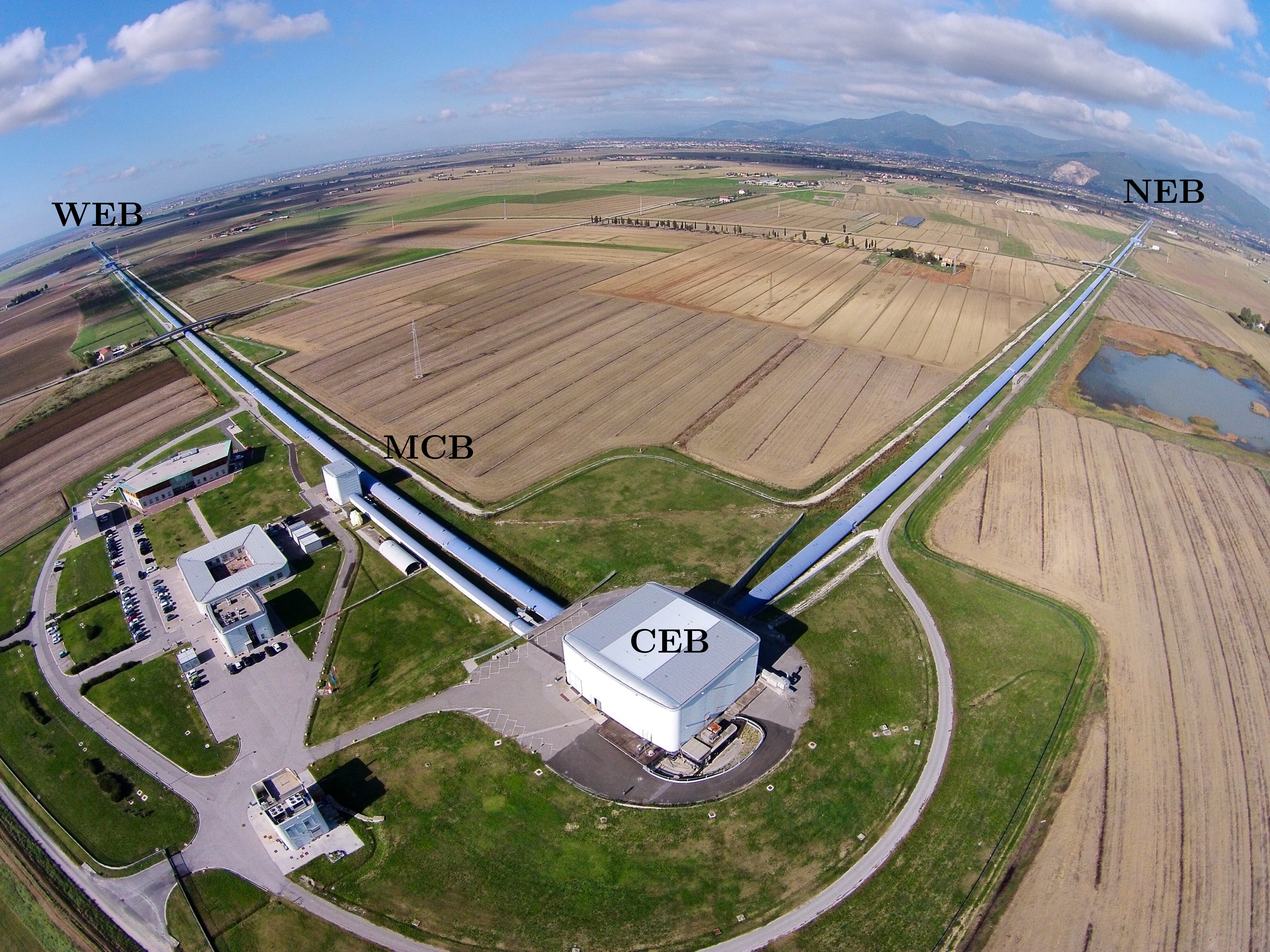}
    \caption{Photography of the Virgo GW detector. The four buildings housing the detector are shown. This study focuses on the NEB and WEB where the end of the arm test masses (mirrors) are installed.}
    \label{fig:Virgo}
\end{figure}

The Virgo interferometer is installed in four buildings called NEB (North End Building), WEB (West End Building), CEB (Central Building), and MCB (Mode Cleaner Building) as can be seen in \cor{Fig. } \ref{fig:Virgo}. The NEB and the WEB are two identical, large buildings ($22\times16.8\times17.2 \ \mathrm{m^3}$), whose walls are not acoustically treated, creating a fairly reverberant acoustic environment.

The test masses (mirrors) are located at the end of the two arms of the interferometer, made up of the vacuum tubes that connect CEB-NEB and CEB-WEB\cor{. They} are subjected to various displacement noises, such as thermal noise, seismic noise,  radiation pressure noise and Newtonian noise \cor{among others} \cite{Maggiore_2007}. Newtonian noise is caused by fluctuations in the gravity field due to mass density variations in \cor{the mediums surrounding} the test masses. These variations can result from vibrations of solids (ground, walls, equipment) and acoustic oscillations both inside and outside the building. To filter out seismic and acoustic noises, the mirrors are decoupled from the
ground using a seven-stage suspension system, called a superattenuator,
installed in a ten-meter high steel tower that acts as a vacuum chamber (residual
pressure = $10^{-6}\ \mathrm{Pa}$) \cite{Basti_2023}. 

Inside the buildings, several equipment (cryogenic pumps, vacuum pumps, cooling fans for electronics) necessary to the detector's operation contribute to the ambient technical acoustic noise. However, the infra-sound acoustic field in these rooms is mainly produced by the HVAC system, which is required to ensure the stability of the ambient temperature ($22\pm0.2\ \mathrm{^\circ  C}$ \cite{Accadia}) and also helps to keep the experimental halls at an overpressure of $7\ \mathrm{Pa}$ to improve air cleanliness. The importance of the impact of the HVAC system on the acoustic noise is highlighted in \cite{Dandrea_TechnicalReportPlanning_2021} where a switch-off test was performed inside the end buildings. The difference between the background acoustic pressure noise and the acoustic pressure when the HVAC is on can reach up to two orders of magnitude in the frequency band 1-150 Hz. Other sources also contribute but they are not included in the
study: environmental sources such as volcanoes, wind, storms, thunderstorms,
and meteors, as well as anthropogenic sources such as wind turbines and
transport vehicles (cars, trains, airplanes, helicopters, motorcycles).

In \cite{Fiori_2020}, a sequential switch off of HVAC systems located in the central building showed a clear impact on the detector's sensitivity, especially in the injection and detection lab where optical elements are directly exposed to the acoustic field. Nowadays, several actions have been undertaken to reduce the HVAC acoustic noise production and coupling with the measurement but this technical equipment remains the main source of acoustic noise at low frequency inside the NEB and WEB.

\medskip

In the terminal buildings, the HVAC system is mainly used for temperature stabilization. The flow rate of the system is controlled manually and is set to the lowest stable working point. This setting is not changed over the course of the year unless important heat waves render the system unable to maintain stable temperature. Those situations can occur mainly during summer. In this study the measurements were taken during spring, in April 2024 when the HVAC was working in its nominal state.

This article focuses on the estimation of Newtonian noise induced specifically by the acoustic field in the NEB and WEB experimental halls caused by the HVAC.  
Newtonian noise is expected to be significant at low frequency (below $10\ \mathrm{Hz}$) \cite{Harms_LowerLimitnewtoniannoise_2022a, Harms_TerrestrialGravityFluctuations_2019,Bader_SeismicnewtonianNoise_2021,Singha_CharacterizationSeismicField_2021}. As a consequence, we limit the frequency band of the study to 1-30 Hz.

\subsection{Acoustic features of the Virgo terminal building}

Among the sources of technical noise, those resulting from the operation of the HVAC system are of prime importance in the low frequency range of interest \cite{Fiori_2020}.

\begin{figure}
    \begin{subfigure}[b]{0.8\textwidth}
        \includegraphics[width=\textwidth]{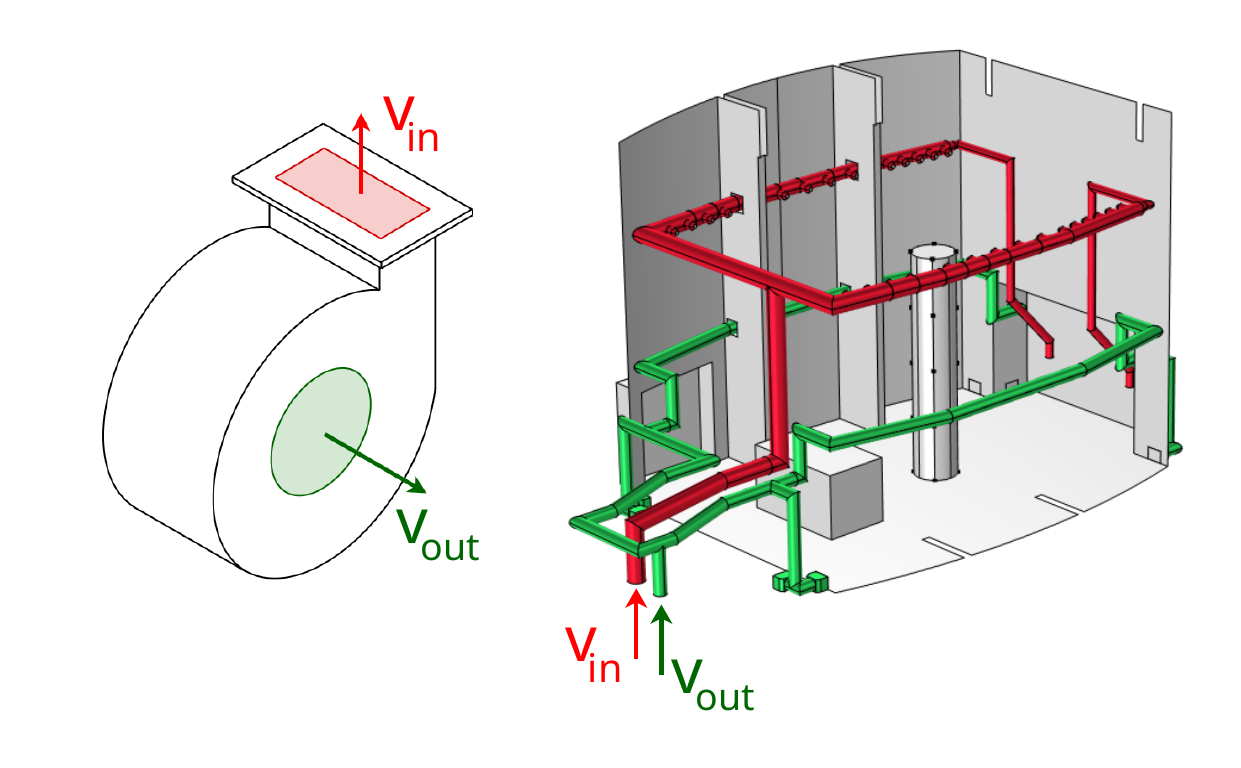}
        \caption{\textit{}}
        \label{fig:3a}
    \end{subfigure}
    \begin{subfigure}[b]{0.8\textwidth}
        \includegraphics[width=\textwidth]{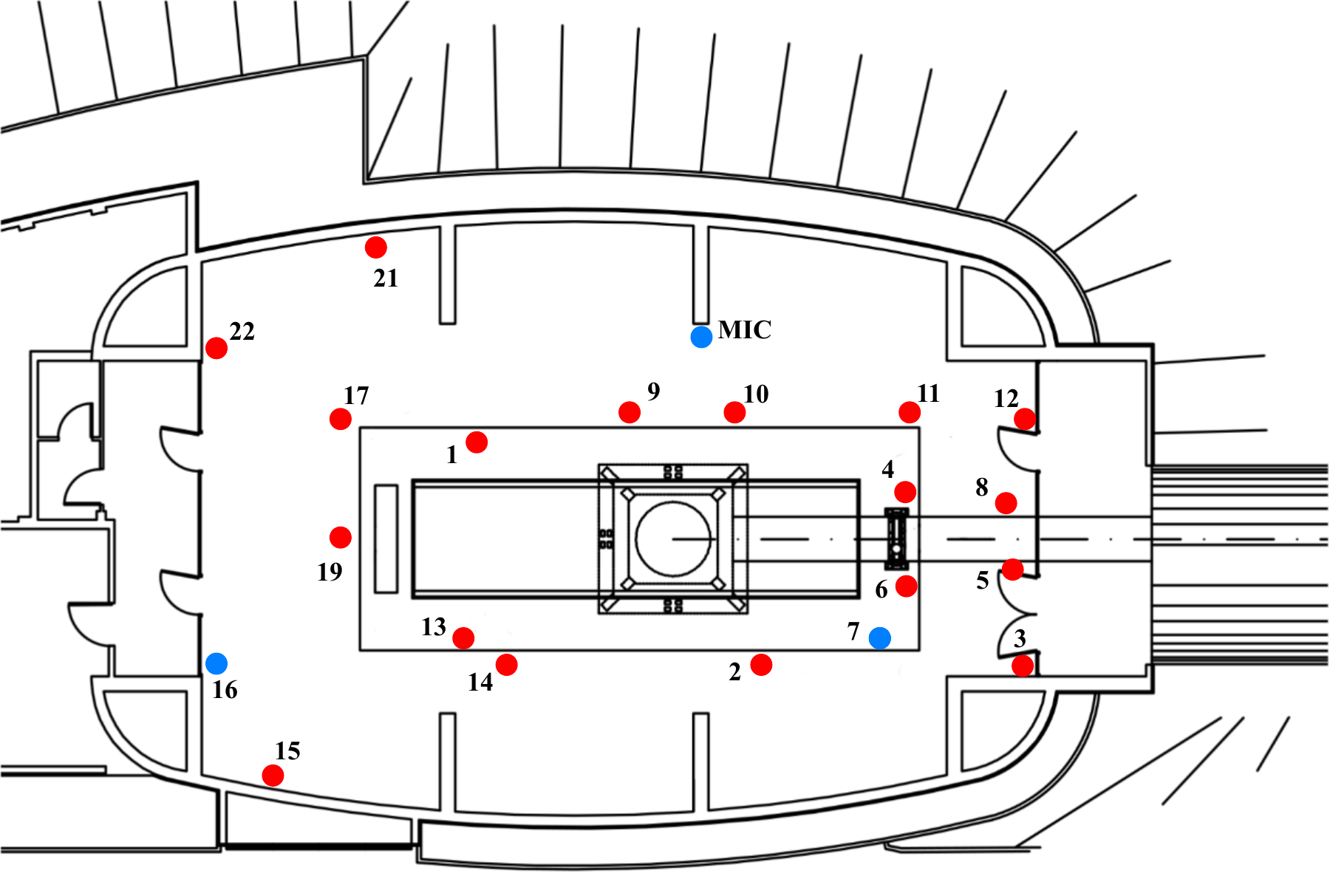}
        \caption{\textit{}}
        \label{fig:3b}
    \end{subfigure}
    \caption{\textbf{(a)} Geometry of the NEB experimental hall considered in the numerical model. The HVAC system ensures air circulation through the suction (green) and injection (red) circuits. \textbf{(b)} Microphone array set up in the WEB. The microphones are located one meter above the ground, only MIC is located at six meters from the ground. Measurements performed on the blue colored microphones are shown in this paper. }
    \label{fig:3}
\end{figure}

The air flow induced by the HVAC propeller is carried into the room through two types of duct network: one network for air suction, (terminated by 6 outlets radiating in the room), represented in green in Fig. \ref{fig:3a}, the other for air injection, (terminated by 32 inlets radiating in the room), represented in red. All inlets and outlets behave as localized acoustic sources that radiate acoustic energy in the room. During the operational time of the detector, environmental conditions within the experimental area remain stable. In the low frequency range, \textit{i.e.} below the Schroeder frequency $f_s$, modal contributions of the acoustic field are resolved and can be identified individually \cite{Kuttruff_2009}. The Schroeder frequency is defined by 

\begin{equation}
    f_s = 2000\sqrt{\frac{RT60}{V}},
\end{equation}

The parameter $RT60$ being the reverberation time of the room in seconds and \textit{V} its volume in $\mathrm{m^3}$. For NEB and WEB, $f_s= 50\ \mathrm{Hz}$.
Above $f_s$, the spacing between two successive eigenfrequency is small enough for modal superposition to lead to a diffuse field that can be described by Sabine's statistical theory. The reverberation time, $RT60$, is also used to estimate the sound absorption of a room using Sabine's law:
\begin{equation}
    RT60 = \frac{0.161V}{\sum_i\alpha_iS_i},
    \label{eq:RT60}
\end{equation}
where $\alpha_i$ is the absorption coefficient of the $i^{th}$ internal surface of the room, named $S_i$, in $\mathrm{m^2}$. In practice, $RT60$ can be measured by performing a noise injection inside the room, usually using an impulse (blanks, balloons) or a high stationary sound level (loudspeaker) that stops abruptly and estimating the time it takes for the acoustic level to decrease by 60 dB. Since the Signal-to-noise ratio (SNR) with a dynamic range of 60 dB is difficult to achieve at low frequencies,  extrapolating results obtained with a decrease of 20 dB ($RT20$) or 30 dB ($RT30$) is usual. The RT30 Reverberation Time per octave band was measured in the NEB at Virgo using balloons to create the excitation and a sonometer \cite{Falxa_2018}. The results are provided in table \ref{tab:1}. The mean absorption coefficients derived from equation \eqref{eq:RT60} are also presented. 

\begin{table}[ht]
\caption{Values of the Reverberation Time RT60 per octave band measured in \cite{Falxa_2018} and associated absorption coefficient $\alpha$ computed using equation \eqref{eq:RT60}.}
\label{tab:1}
\hrule 
\vspace{0.2cm}
\resizebox{\textwidth}{!}{
\begin{tabular}{l|ccccccc}
     $f$ (Hz) & 125 & 250 & 500 & 1000 & 2000 & 4000 & 8000\\
     $RT60\ \mathrm{(s)}$ & 5.22 & 5.44 &  5.35 &  4.79 & 4.46 & 3.86 & 3.86\\
     $\alpha$ & 0.08 & 0.08 & 0.08 & 0.09 & 0.1 & 0.11 & 0.11\\
\end{tabular}}
\vspace{0.2cm}
\hrule 
\end{table}

In the frequency band of interest, 1-30 Hz, the absorption coefficient is, therefore, assumed to be equal to $\alpha=0.08$, which is the value obtained for the lowest available octave band (125 Hz). This value is possibly overestimated since the Sabine theory is valid only if the field is diffuse \textit{i.e.} above $f_s=50\ \mathrm{Hz}$. Nonetheless, this assumption remains a first attempt at incorporating relevant losses into the numerical model.

In harmonic regime, the acoustic field resulting from a set of sources can be described using a sum of modal contributions associated with each room mode. The number of room modes to consider is infinite. In practice, for the simulation, a truncated modal basis is used, where the eigenfrequency of the highest order mode taken into account is at least twice the maximum frequency of the band of interest. Due to the volume of the room, we expect the number of modes to be high. It is described by the mode count function, which value at the frequency $f$ is the number of modes below $f$. For a room, it is given by \cite{Kuttruff_2009}

\begin{equation}
    N = \frac{4\pi}{3}V\left(\frac{f}{c}\right)^3 +\frac{\pi}{4}S\left(\frac{f}{c}\right)^2+\frac{Lf}{8c},
\end{equation}

where $V$ is the volume of the room in $\mathrm{m^3}$, $S$ the total surface area of the walls, floor and ceiling in $\mathrm{m^2}$, $L$ the total length of all edges in $\mathrm{m}$, $c$ the \cor{speed of sound} in $\mathrm{m.s^{-1}}$. For NEB geometry, using $V = 5000\ \mathrm{m^3}$, $S = 1740\ \mathrm{m^2}$ and $L = 206\ \mathrm{m}$, the mode count at $f=30\ \mathrm{Hz}$ is $N(30\ \mathrm{Hz})=50$, which is high even at such low frequencies. The numerical study performed thus involves 88 modes that are used to reconstruct the acoustic pressure field in the frequency range 1-30 Hz.

\subsection{Methodology for Gravity field computation}

In a fluid at rest, in the linear regime, the adiabatic acoustic pressure fluctuations $\delta p$ are proportional to the acoustic fluctuations of the mass density $\delta \rho$:

\begin{equation}
    \delta \rho(\vec{r},t) = \frac{\rho_0}{\gamma p_0} \delta p(\vec{r},t),
    \label{adiabatic}
\end{equation}

with $\rho_0$ the air density in $\mathrm{kg.m^{-3}}$, $p_0$ the static pressure in $\mathrm{Pa}$, $\gamma$ the heat capacity ratio and where $\vec{r}$ is the position vector and $t$ is time. The mass distribution around an observation point can therefore evolve over time, giving rise to fluctuations in the gravity field, known as Newtonian noise. The background on the mathematical framework used to model the Newtonian noise level of acoustic origin is provided in \cite{Fiorucci_2018}. The variation of the gravity field $\delta \vec{g}(\vec{r},t)$ is derived from the gradient of the gravitational potential $\delta \Phi(\vec{r},t)$ as follows:

\begin{equation}
    \vec{\delta g}(\vec{r},t) = - \vec{\nabla} \delta \Phi(\vec{r},t),
    \label{g}
\end{equation}

The fluctuations of the gravitational potential $\delta \Phi(\vec{r}, \omega)$ are obtained from Newton's law of gravitation :

\begin{equation}
    \delta \Phi(\vec{r}, \omega) = G \int_V \frac{\delta \rho (\vec{r'}, \omega)}{|\vec{r'} - \vec{r}|} dV(\vec{r'}),
    \label{dphi}
\end{equation}

where $G$ is the gravity constant in $\mathrm{N.m^2.kg^{-2}}$ and $V$ is the volume of the room that defines the integration domain. The acoustic pressure field in the room is described as a modal expansion \cite{Bruneau_Scelo_2006,Fahy_Gardonio_2007}:

\begin{equation}
    \delta p(\vec{r}, \omega) = \sum_n
    A_n(\omega) \phi_n(\vec{r}),
    \label{eq:intmodal}
\end{equation}

where $A_n$ are the modal amplitudes and $\phi_n$ the acoustic eigenshapes, the subscript $n$ indicates the modal indices. For a set of monopolar sources located inside the room, the resulting modal amplitudes are given by \cite{Bruneau_Scelo_2006}:

\begin{equation}
    A_n(\omega) = jk\rho_0c\sum_i^{N_s}\frac{\phi_n(\vec{r_i})Q_i(\omega)}{\Lambda_n(k_n^2-k^2)},
    \label{eq:An}
\end{equation}

with $Q_i$ the acoustic flow of the $i^{th}$ monopolar source in $\mathrm{m^3.s^{-1}}$, $\vec{r_i}$ its position vector, $k_n$ is the $n^{th}$ mode in $\mathrm{m^{-1}}$, $k=\omega c^{-1}$ is the wavenumber in $\mathrm{m^{-1}}$ and $\Lambda_n$ is the mode's norm. $N_s$ is the number of sources to be considered. $\Lambda_n$ is defined as 
\cor{ 
\begin{equation}
   \Lambda_{n}  = \int \phi_n\phi_n^*dV, 
\end{equation}}

By substituting equation \eqref{adiabatic} in the expression of $\delta \Phi$ as given by equation \eqref{dphi}, one obtains $\delta \Phi$ as a function of $ \delta p$. Writing $\delta p$ as a modal sum, the fluctuations of the gravity potential $\delta \Phi$ are finally expressed as follows:
\begin{equation}
    \delta \Phi(\vec{r}, \omega) = -\sum_{n}
    B_n(\omega) \psi_n(\vec{r}),
\end{equation}
where
\begin{equation}
    B_n(\omega) = - A_n(\omega) \frac{G \rho_0}{\gamma p_0},
    \label{Bn}
\end{equation}
and
\begin{equation}
    \psi_n(\vec{r}) = \int_V \frac{\phi_n(\vec{r'})}{|\vec{r'} - \vec{r}|} dV(\vec{r'}),
    \label{psin}
\end{equation}
so that
\begin{equation}
    \delta \vec{g}(\vec{r},\omega) = \sum_{n}
    B_n(\omega)  \vec{\nabla} \psi_n(\vec{r}),
    \label{deltag}
\end{equation}
As a consequence, fluctuations of the gravity field are described in equation \eqref{deltag} as an expansion over a functional basis $\nabla\psi_n$, whose elements $\psi_n$ are explicitly derived from the acoustic eigenshapes $\phi_n$. Lastly, the strain noise analogous to the measured quantity of the detector is obtained as follows:

\begin{equation}
    h(\omega)=\frac{2{\delta g_x}(\omega)}{L\omega^2},  
\end{equation} 

where $g_x(\omega)=\delta\vec{g}(\vec{r},\omega).\vec{x}$ is the axial component of the gravity field aligned with the arm of the interferometer. $L$ is the arm's length in m (in the case of Virgo, $L=3000\ \mathrm{m}$). The factor 2 comes from the uncorrelated summation of the 4 test masses contribution \cite{Maggiore_2007}. In this article, we assume that the test mass is not subject to any external forces. However, in reality, the test mass is suspended from a pendulum chain (the superattenuator), which affects its motion, particularly at low frequencies. Accounting for the dynamics of the pendulum chain is therefore necessary to improve the accuracy of the estimate at frequencies below a few hertz.

\subsection{Numerical implementation using the finite element method}

The numerical modeling is based on the finite element method as implemented in the commercial software Comsol Multiphysics and more specifically in the Acoustic module. The study is performed in the frequency domain only.

\medskip

Two models called full model and simplified model are considered. The full model includes both the duct networks and the room. Its geometry is depicted in Fig. \ref{fig:3a}. The mesh is made of tetrahedral elements whose maximum size is chosen according to the 6 elements per wavelength rule. At 30 Hz, the acoustic wavelength is 11.8 m meaning that the maximum size of an element is 1.69 m. Due to the duct curvature, the minimum element size is set to 0.123 m. The shape and basis function are both Lagrange quadratic functions. In this iteration of the model, the mesh is made of 889886 elements. This configuration is used to define the full model and is used to estimate the $Q_i$ acoustic flow used in equation \eqref{eq:An}.

\medskip

The simplified model does not include the ducts. The mesh follows the same rules as the full model and is composed of 84232 elements. This simplified model is used to compute the eigenmodes and eigenfrequencies of the room and the associated gravity potential functions.

\medskip

Some geometrical details are omitted due to the low frequency range of the study to keep the number of elements required for the mesh low. The convective effects are not taken into account due to the low mach number of the air flow ($M \approx 0.02$), furthermore, the temperature gradients of the field are not considered. The walls are supposed rigid since vibroacoustic coupling is ignored and no structure-borne excitation is taken into account. This assumption is due to the fact that walls of the experimental area are made of concrete with high mass density and stiffness rendering the coupling with the acoustic field low. Boundary conditions considered during the simulation process are of two sorts, Neumann boundary conditions to enforce the rigid wall hypothesis and impedance boundary condition to insert the $\alpha = 0.08$ absorption factor estimated in section \ref{sec:2-NNmain}. The two boundary conditions are applied on all boundaries of the domain.

\section{Results\label{sec:3-Results}}

Computing the gravity field resulting from pressure fluctuations requires to \cor{model} the acoustic \cor{field in the room}. Acoustic measurements are first presented to validate the numerical model. The induced gravitational field is then calculated through direct integration and presented as a series expanded in functional form. 

\subsection{Acoustic}

\subsubsection{\cor{Estimation of the equivalent source for the HVAC system}}

\cor{The full model described in section \ref{sec:2-NNmain} is used to estimate the equivalent source for the HVAC. The HVAC is described in the model as pistons acting at the entrance of the suction and injection ducts. The velocities $V_{in}$ and $V_{out}$ (see Fig. \ref{fig:3}) are unknown and should be fitted to the measurement. In a first stage, we compute the acoustic field assuming an imposed unitary normal velocity $V=V_{in}=V_{out}=1 \ \mathrm{m.s^{-1}}$. The relative phase between the two pistons is thus supposed to be the same. The acoustic field simulated in this condition provides directly the transfer function $H_n=P_n.V^{-1}$ with $P_n$ the pressure at the location of the $n^{th}$ microphone and V the piston velocity. For this simulation, a direct solver is used to compute the solution (MUMPS).
From the measurements, one can estimate the power spectral density of the piston from:}

\begin{equation}
    S_{pn} = |H_n|^2S_V    
\end{equation}

\cor{where $S_{pn}$ and $S_V$ are the power spectral densities of the $n^{th}$ microphone and the acoustic piston, respectively, in $\mathrm{Pa^2.Hz^{-1}}$ and $\mathrm{m^2.s^{-2}.Hz^{-1}}$. The location of said microphones is shown in Fig. \ref{fig:3b}. Assuming a uniform power spectral density over the entire frequency band (1-30 Hz), the piston power spectral density that minimizes the error on the 22 microphones leads to $S_V=(3.10^{-3})^2 \ \mathrm{m^2.s^{-2}.Hz^{-1}}$. \corsec{More information on the experimental setup used are given in appendix \ref{sec-5A}.} This factor can be interpreted as the square of the velocity of a piston equivalent to the HVAC. This factor is used to scale all the simulated results.} \corsec{This estimate is evaluated using experimental data taken on April 13 2024 between 00:00:00 and 01:00:00 UTC. As the estimate directly depends on the amplitude of the acoustic noise, yearly evolution of this noise is shown in appendix \ref{sec-5B}.}

\begin{figure}
    \begin{subfigure}[b]{\linewidth}
        \includegraphics[width=\linewidth]{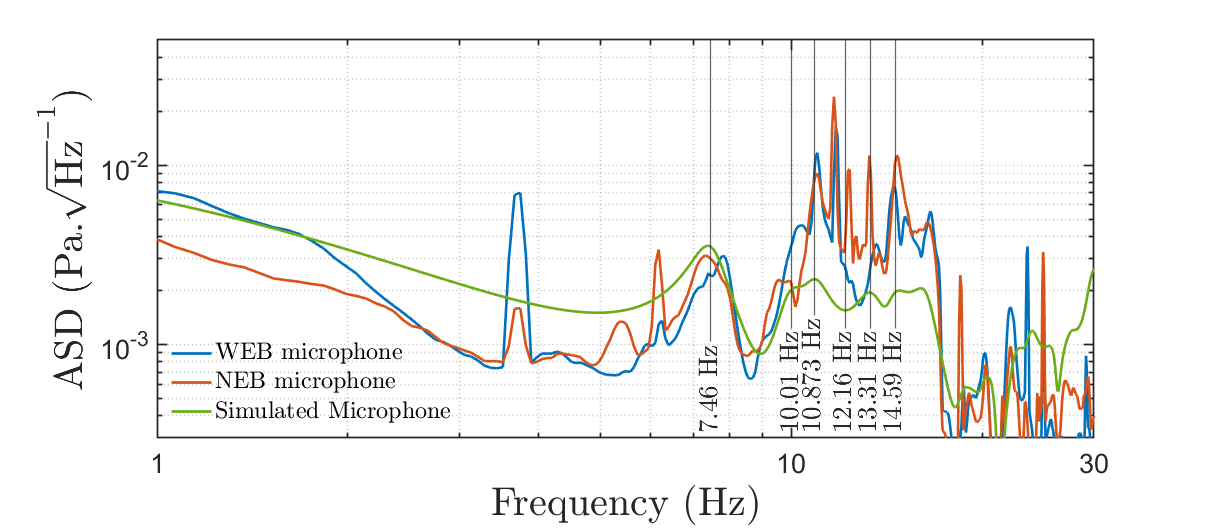}
        \label{fig:validateENV}
        \caption{}
    \end{subfigure}
    \begin{subfigure}[b]{\linewidth}
        \includegraphics[width=\linewidth]{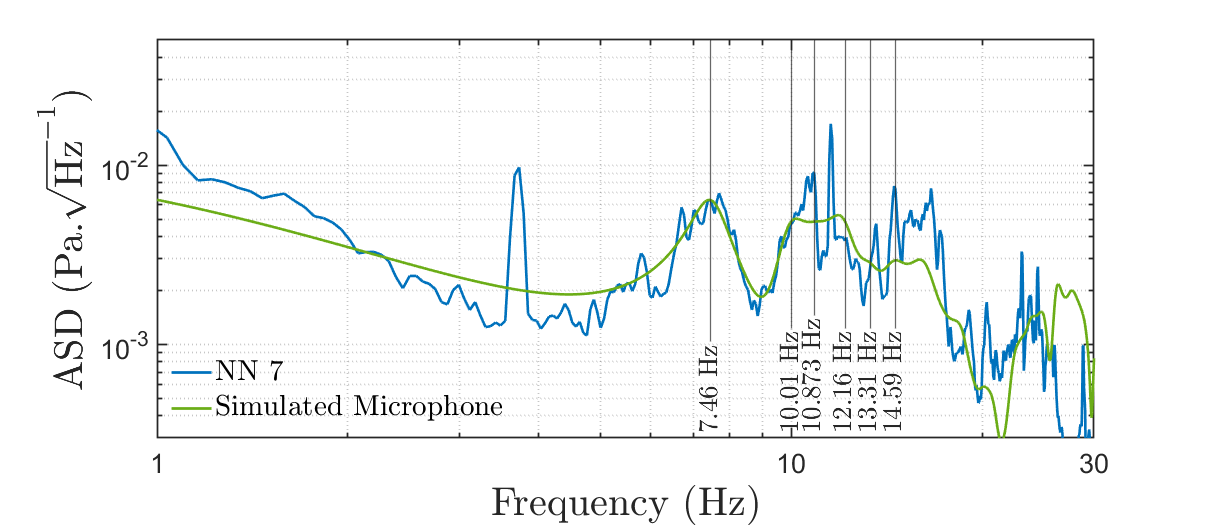}
        \label{fig:validateNN7}
        \caption{}
    \end{subfigure}
    \begin{subfigure}[b]{\linewidth}
        \includegraphics[width=\linewidth]{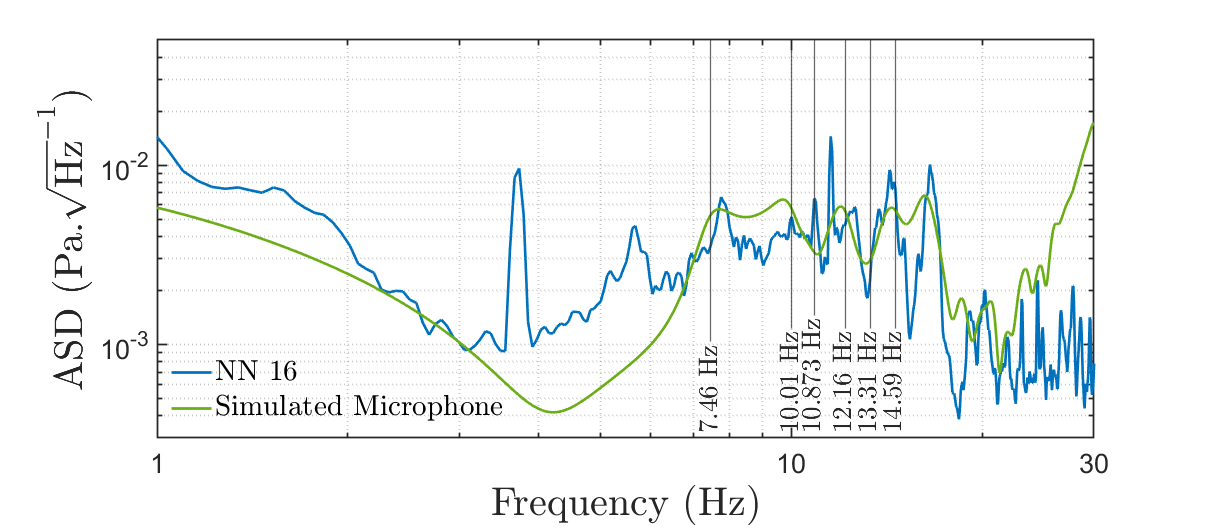}
        \label{fig:validateNN16}
        \caption{}
    \end{subfigure}
    \caption{\textbf{(a)} Acoustic pressure measured in the NEB and the WEB at the MIC location (As defined by the blue point in Fig. \ref{fig:3}). \corsec{Data were extracted on April 13 2024 between 00:00:00 and 01:00:00 UTC. Data are sampled at 500 Hz. 10 seconds long Hann windows with 50\% overlap are used for computation of the ASD.} The green curve is the simulated pressure at the same position. \textbf{(b)} Measured and simulated pressure at microphone 7. \textbf{(c)} Measured and simulated pressure at microphone 16.}
    \label{fig:5}
\end{figure}

\subsubsection{Validation of the acoustic model}

\cor{Fig. \ref{fig:5} shows a comparison between the modeled acoustic field and measurements performed in the NEB and the WEB (orange and blue curves). The spectra of the measurements from the two end buildings are very similar, demonstrating that the two rooms have a twin-like behavior in terms of acoustics. The acoustic eigenfrequencies are indicated using vertical lines. As the modal overlap is strong, the modes are not well resolved.}

\cor{Qualitative comparison between the microphone array spectra available in the WEB and the simulated ones are made in the following paragraph.} At low frequencies, a slope difference can be observed. This can be explained by two main hypotheses: The model is perfectly airtight. In this configuration, a 0 Hz resonance is a solution to the wave equation, and the ultra-low frequency behavior is then controlled by this "uniform pressure" mode. In reality, air leakage in the room dampens the resonances of the building and generates a low frequency Helmholtz type mode. It is usually strongly damped and remains outside the measurement range of most microphones but dictates the slope of the spectrum at such low frequencies. The second hypothesis is the source approximation. As described by \cite{Jiang,neiseNoiseReductionCentrifugal1976}, aeroacoustic sources can be decomposed into two different components. The first is tonal and is tied to the characteristic frequencies of several parts of the system. In Fig. \ref{fig:5}, several sharp components can be identified, some of them have been linked to sub-assemblies of the HVAC system  (gears, motor and belt of the reducing gear group). Another important tonal contribution comes from the fan's rotation and from the blade passing frequency being a multiple of the rotation speed by the number of blades of the rotor. The second component is a broadband contribution generated by the vorticity of the air flow. The frequency range of this contribution varies greatly with the geometry of the system and its working point. A general rule of thumb for this noise is that its amplitude follows a power law of the rotation speed. \cor{In the simulation, the tonal components are not taken into account and the acoustic source is described by an approximate acoustic piston whose amplitude is uniform over the frequency band of the study.}

\subsubsection{Simplified acoustic model}

Even with these discrepancies, the model's quadratic error compared with measurements averaged over the entire microphone antenna lies at 9.8\%. For microphones located close to some equipment not taken into account by the model, this error increases due to not described scattering phenomena. We consider the accuracy to be sufficient to correctly represent the acoustic field inside the terminal buildings. This computation is heavy (between 12 and 24 hours of computational time depending on the mesh quality), and the complexity of the duct geometry renders parametric studies complicated.

\medskip

\begin{figure*}
    \begin{subfigure}[b]{0.49\linewidth}
        \includegraphics[width=\linewidth]{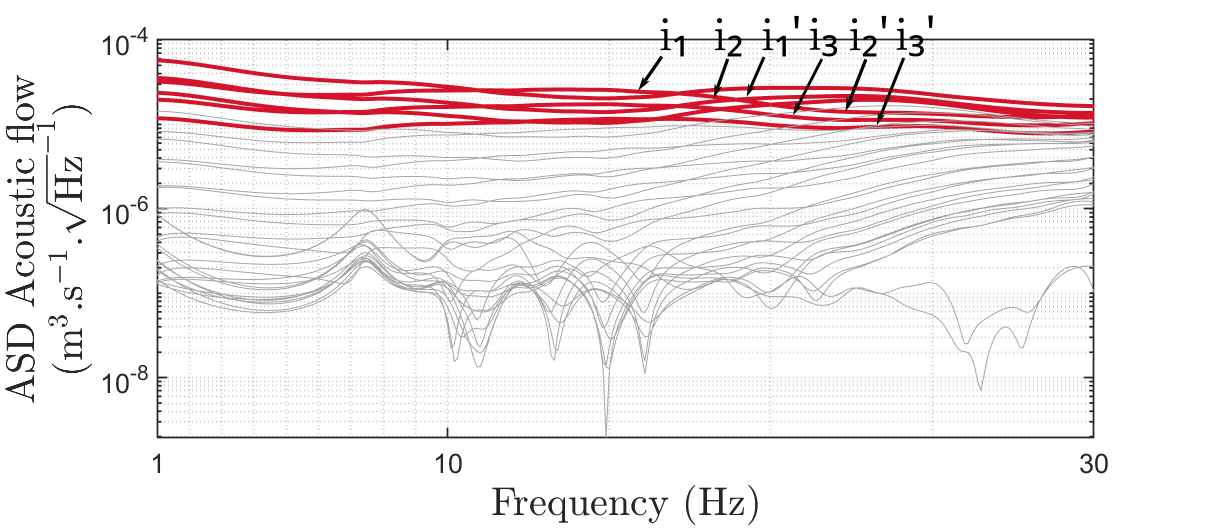}
        \caption{{}}
        \label{fig:5a}
    \end{subfigure}
        \begin{subfigure}[b]{0.49\linewidth}
        \includegraphics[width=\linewidth]{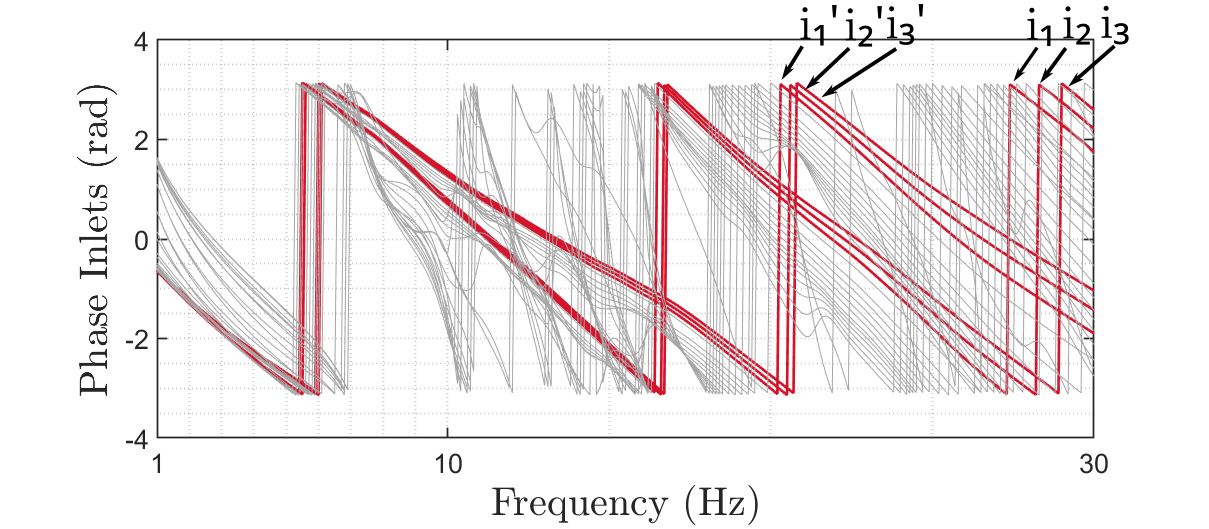}
        \caption{{}}
        \label{fig:5b}
    \end{subfigure}
    \begin{subfigure}[b]{0.49\linewidth}
        \includegraphics[width=\linewidth]{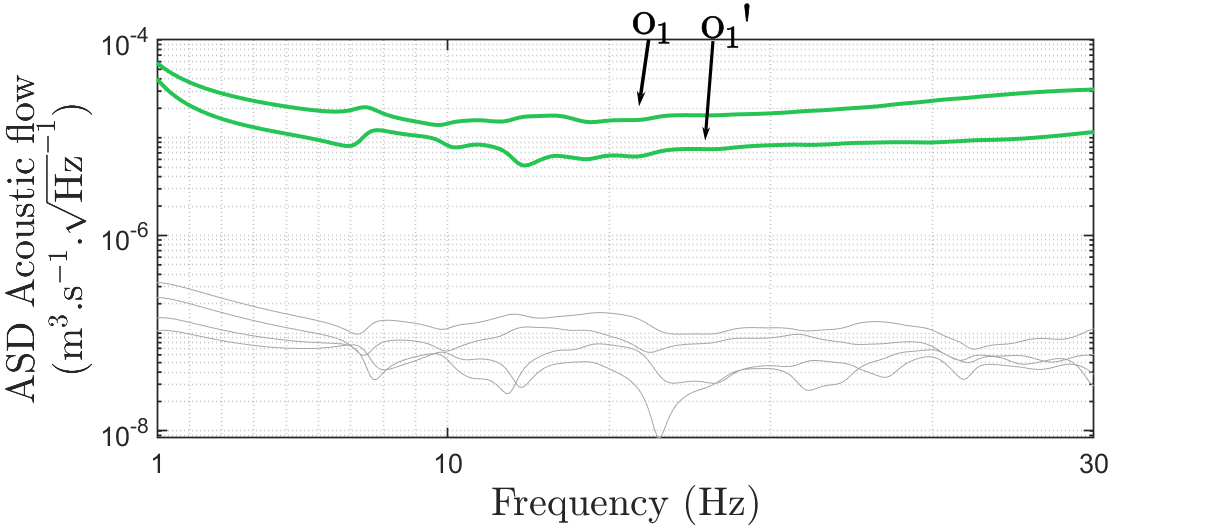}
        \caption{}
        \label{fig:5c}
    \end{subfigure}
        \begin{subfigure}[b]{0.49\linewidth}
        \includegraphics[width=\linewidth]{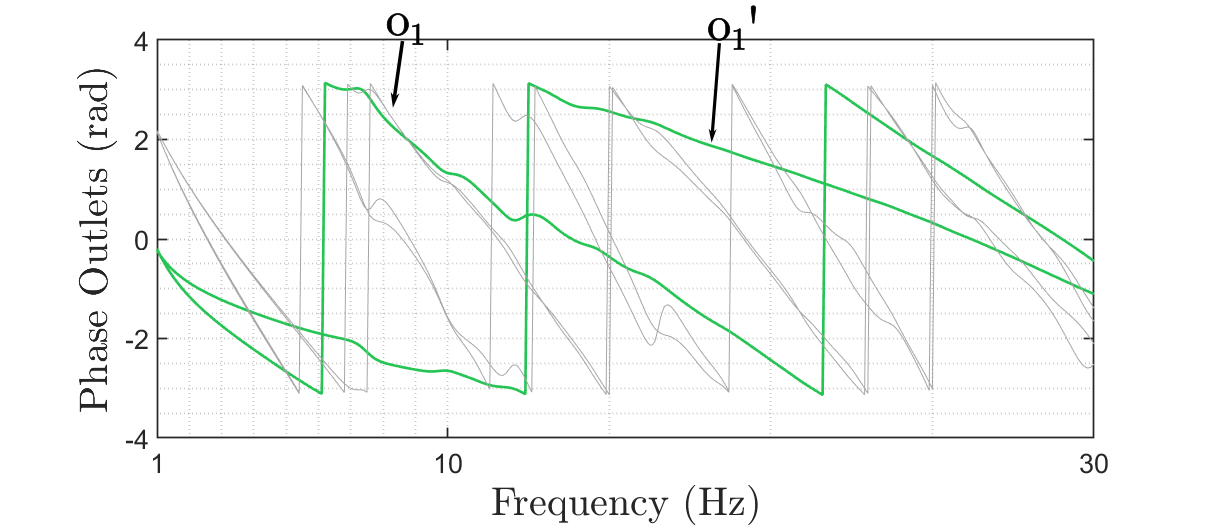}
        \caption{{}}
        \label{fig:5d}
    \end{subfigure}
    \begin{subfigure}[b]{0.24\linewidth}
        \includegraphics[width = \linewidth]{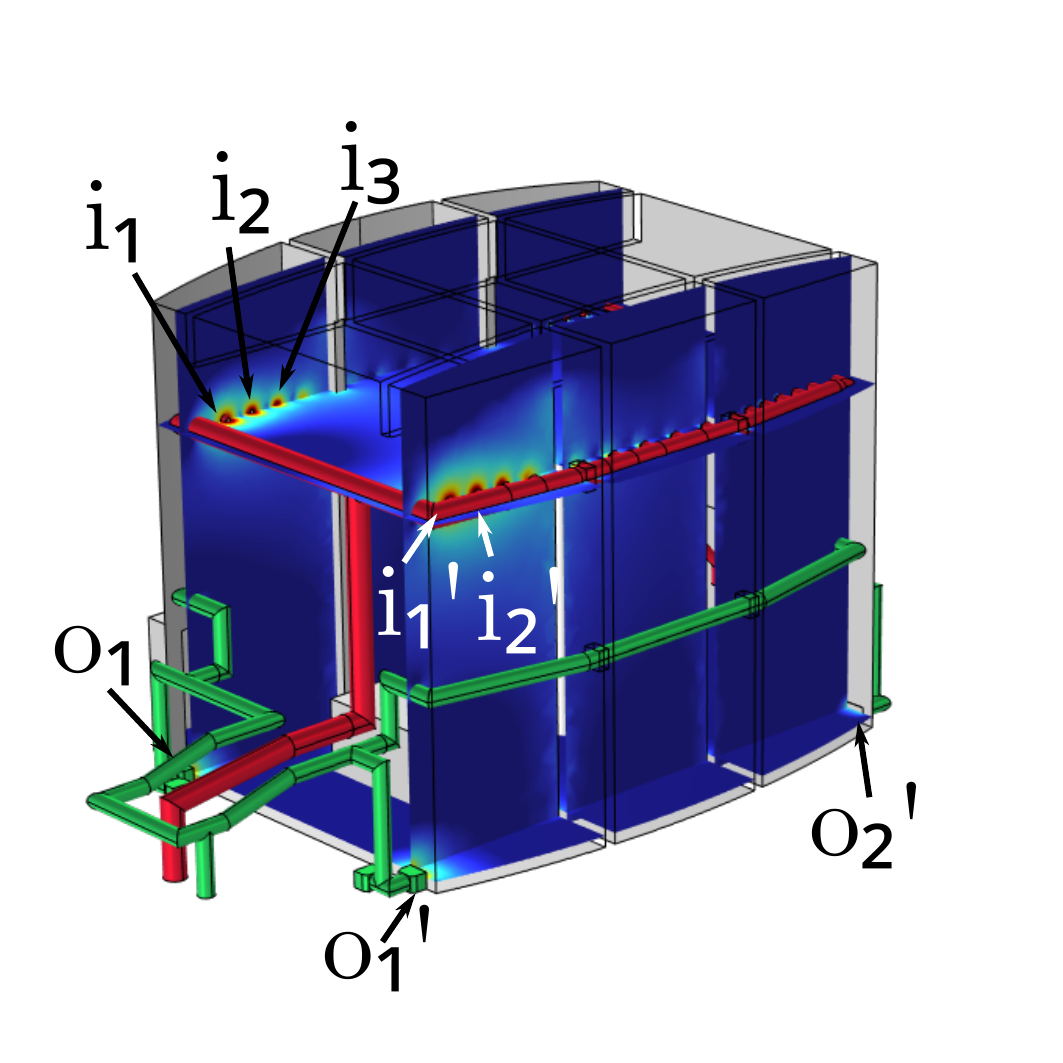}
        \caption{}
        \label{fig:5e}
    \end{subfigure}
    \begin{subfigure}[b]{0.24\linewidth}
        \includegraphics[width = \linewidth]{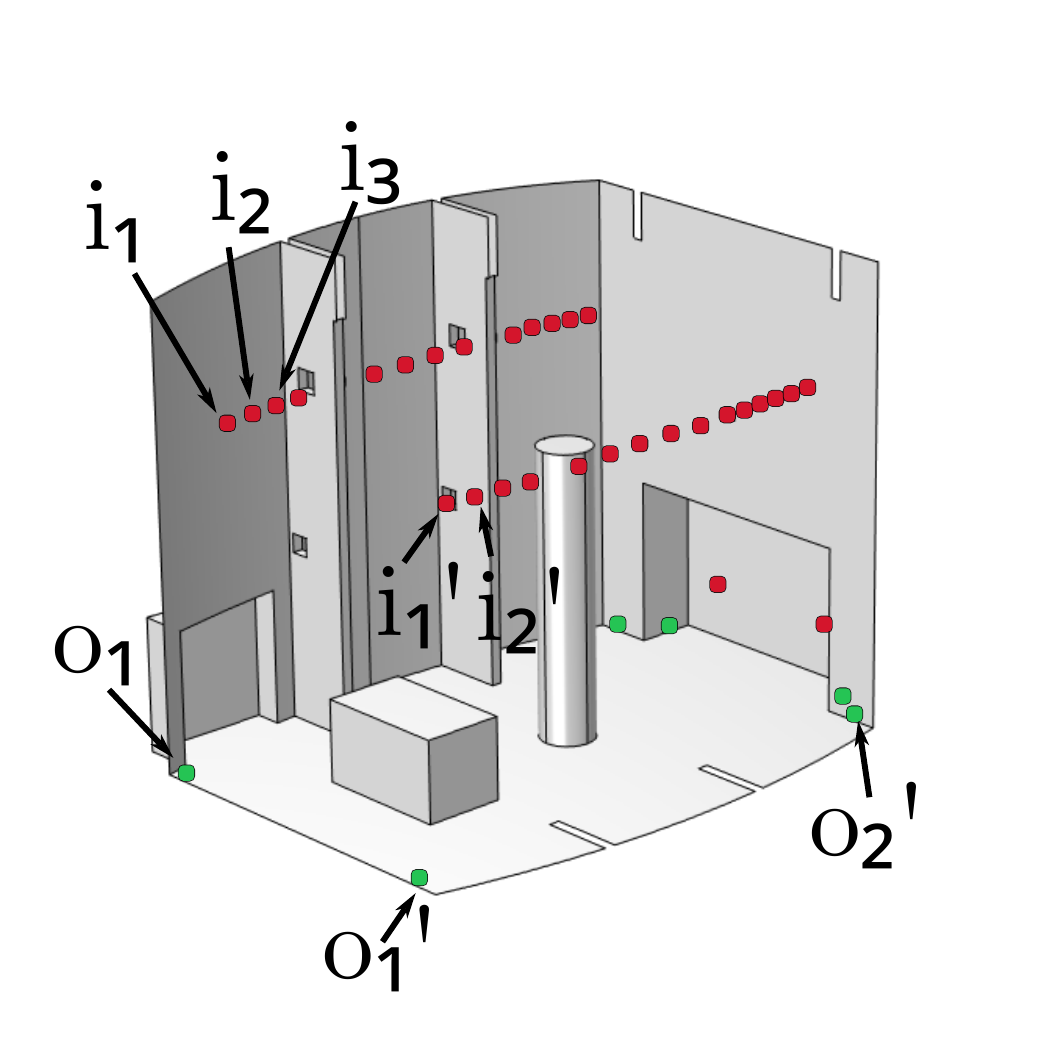}
        \caption{\textit{}}
        \label{fig:5f}
    \end{subfigure}
    \hfill
    \begin{subfigure}[b]{0.49\linewidth}
        \includegraphics[width=\textwidth]{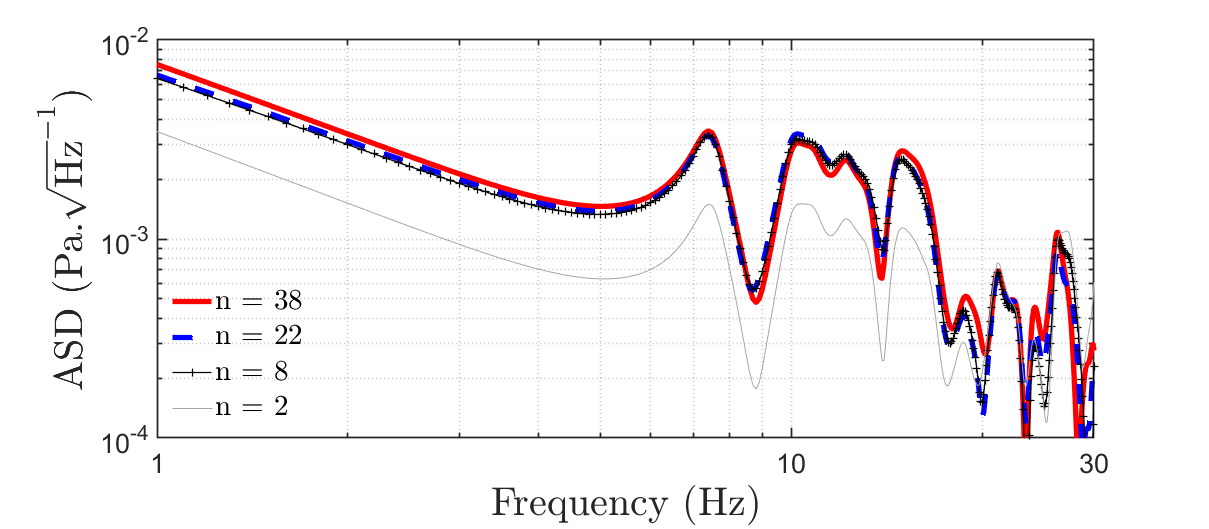}
        \caption{}
        \label{fig:5g}
    \end{subfigure}
    \caption{\textbf{(a)} Normal acoustic flow computed for each inlets. \textbf{(b)} Phase of each inlets. \textbf{(c)} Normal acoustic flow computed for each outlets. \textbf{(d)} Phase of each outlets. \textbf{(e)} Naming convention of each inlets and outlets for the full model. The cut plane are displaying the acoustic velocity. \textbf{(f)} Naming convention of the simplified model. \textbf{(g)} Monopoles number variation effect on the spectrum at the MIC microphone's location.}
    \label{fig:verif}
\end{figure*}

To facilitate this procedure, a simplification of the numerical setup is proposed. This process is described in Fig. \ref{fig:verif}. \cor{Taking advantage of the large wavelength to diameter ratio, the ducts are replaced by a set of acoustic monopoles whose acoustic flow $Q_i$ is estimated using the full model.}


\cor{The full model is used to calculate the transfer function between the acoustic velocity at the center of each inlet or outlet and the equivalent piston velocity in the HVAC system. Multiplying this transfer function by the piston velocity $\sqrt{S_V}$ yields the amplitude and phase of the velocity $\tilde{V_{i}}$ at the location of each inlet and outlet $i$. This quantity is called the amplitude spectral density (ASD). Since the plane wave approximation holds in the duct, the acoustic velocity $\tilde{V_{i}}$ is quasi uniform across the surfaces of the inlets and outlets. The acoustic flow rate $Q_i$ of an inlet or outlet $i$ corresponds to the flow rate passing through the surface of the disk terminating the duct. It also corresponds to the flow rate passing through the surface of a sphere centered at that end. We can therefore write $Q_i=\pi r^2 \tilde{V}_{i}=4\pi r^2 V_i$, where $V_i$ is the acoustic velocity of a pulsating sphere of radius $r$. Consequently, $V_i = \tilde{V_{i}}/4$ and therefore}

\begin{equation}
    Q_i = \frac{\tilde{V_{i}}}{4} \pi r^2  
\end{equation}

 \cor{The computed acoustic flows for the monopoles representing the inlets and outlets are respectively shown in Fig. \ref{fig:5a} and Fig. \ref{fig:5c}. The corresponding phases for the inlets and outlets are shown in Fig. \ref{fig:5b} and Fig. \ref{fig:5d}}. Fig. \ref{fig:5e} shows a view of the full model, where vertical cutting planes are overlaid on the geometry. The color represents the modulus of the acoustic velocity. The colored regions reveal the acoustic radiation of the inlets and outlets inside the room.

\medskip

\cor{Using the simplified model, we compute the acoustic eigenmodes $\phi_n$ and eigenfrequencies $\omega_n$. Because of the wall acoustic absorption described by the imposed impedance boundary condition, these quantities are found to be complex. As a consequence the modal wave number $k_n=\omega_n.c^{-1}$ in equation \ref{eq:intmodal} is complex. The modal damping coefficient $\chi_n$ is therefore given by $\chi_n = \Im (\omega_n)/\Re (\omega_n)$. The solver ARPACK is used to solve the eigenvalue problem. The mode's norm $\Lambda_n$ is also computed in the software.} The integro-modal method described in section \ref{sec:2-NNmain} is then used to compute the response of the room \cor{using MATLAB R2024b}. The reconstructed acoustic field using formula \ref{eq:intmodal} is depicted in Fig. \ref{fig:5e}. In this figure, we perform a parametric study on the number of sources included in the model. Each curve represents the field for a set number of monopoles, ranging from the full set of 38 to only 2. The sources are deleted 2 by 2 starting from the end of the duct networks. With only the first 6 inlets and 2 outlets considered, the mean quadratic error remains below 15\%. The following computation will therefore only include the 8 aforementioned sources. This simplified model, when compared to the full simulation, displays a higher average error, however, the computation of the acoustic field becomes straightforward using the field equation shown in section \ref{sec:2-NNmain}.

\subsection{Gravity}

\begin{figure*}
    \begin{subfigure}[b]{\textwidth}
        \centering
        \includegraphics[width = \textwidth]{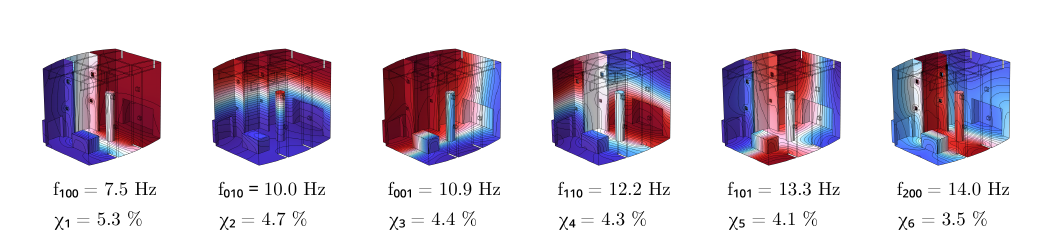}
        \caption{$\phi_n$}
        \label{fig:Phi}
    \end{subfigure}
    \hrule
    \begin{subfigure}[b]{\textwidth}
        \centering
        \includegraphics[width = \textwidth]{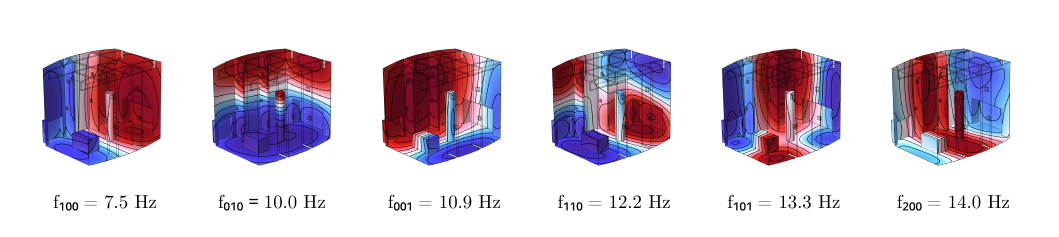}
        \caption{$\psi_n$}
        \label{fig:Psi}
    \end{subfigure}
    \hrule
    \begin{subfigure}[b]{\textwidth}
        \centering
        \includegraphics[width = \textwidth]{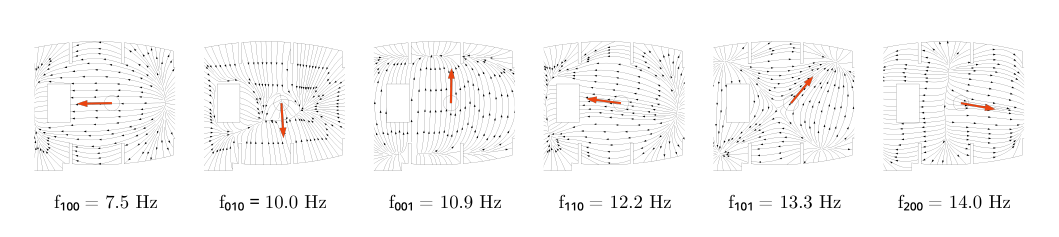}
        \caption{$\vec \nabla\psi_n$}
        \label{fig:NablaPsi}
    \end{subfigure}
    \hrule
    \begin{subfigure}[b]{\textwidth}
        \centering
        \includegraphics[width = \textwidth]{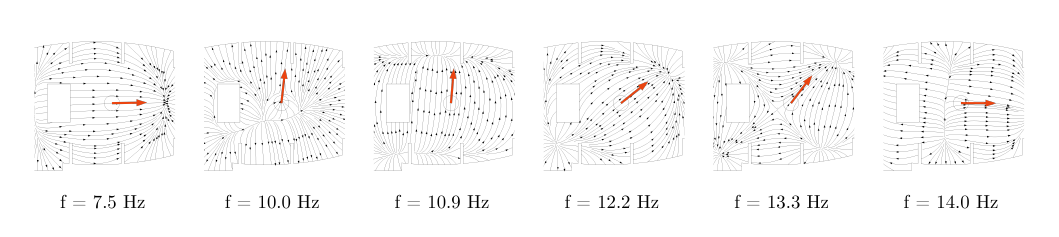}
        \caption{$\vec{g}(f)$}
        \label{fig:g}
    \end{subfigure}
    \caption{\textbf{(a)} Firsts 8 acoustic eigenfrequencies and eigenshapes of the NEB $\phi_n$. $\chi$ indicates the modal loss factor. \textbf{(b)} Gravity potential field function $\psi_n$ of the room. \textbf{(c)} Streamlines for the gravity field \corsec{function $\nabla\psi$} in the room in the 2D cross-section at one meter from the ground. \textbf{(d)} Streamlines of the acceleration field in the 2D cross-section at 1 meter from the ground. This field is the acceleration generated through gravitational attraction by the acoustic pressure field when it is excited by the HVAC system.}
    \label{fig:acoustic}
\end{figure*}

Once the acoustic field is known, we now focus on estimating the gravity field it generates. As stated in section \ref{sec:2-NNmain}, variations of the acoustic pressure translate into a local variation of the air's mass density which in turn creates noise on the detector signal through gravitational attraction of the test masses. Following the mathematical framework presented earlier, we can compute the functional basis associated with the acoustic eigenshapes and estimate the modal amplitudes for each mode. \cor{The functionals associated with the potential and gravity fields are computed using the gauss quadrature method implemented in the software. A $4^{th}$ order quadrature is giving accurate results. The gradient operation is then again performed in the software.} The gravity field can then be recovered by summation over each acoustic mode's response to the set of sources. Fig. \ref{fig:acoustic} shows in each of its row a step in the computation of the gravity field. The first row shows the first six acoustic eigenshapes computed considering only the experimental area, along with their eigenfrequencies and the estimated modal loss factor computed from the absorption coefficient discussed in the last paragraph. The eigenvalue naming convention follows a classical modal naming scheme, where the three indices used represent the order of the solution for each dimension of the domain. \cor{The referential used is cartesian. In this case, the x axis is aligned with the arm of the interferometer, the y axis along the height of the building and the z along the width.} The first number is associated to the x axis, the second to the y axis and the last to the z axis. An eigenshape having only one non zero digit is called an axial mode, one with two indices is a tangential mode and one with the three non-zero digit is called an oblique mode. The second row shows the gravity potential \corsec{functions} associated to the acoustic eigenshapes computed using equation \eqref{psin}. The third row displays the projection of the functional basis in the xy plane at the level of the mirror. The red arrow indicates the direction of the field at the location of the mirror, its amplitude does not give any information regarding the strength of the acceleration. Finally, using equation \eqref{deltag}, we can compute the response of the gravity field to acoustic stimuli. It is displayed in the $4^{th}$ row of the figure. The mirror's radius of curvature is large ($\sim 1.5\ \mathrm{km}$), we assume, for the sake of simplicity, that the mirror's surface is flat. In this approximation, and by considering no tilt of the mirror, the only component of the displacement to which the interferometer is sensitive is the longitudinal motion in the direction of the laser beam. The vectors are defined in cartesian coordinates, with the x coordinate being in the direction of the laser beam. As such, only the x component of the gravitational field vector is of interest for the study. 

\subsection{Parametric study}

\begin{figure*}
    \begin{subfigure}[b]{0.24\linewidth}
        \includegraphics[width = \linewidth]{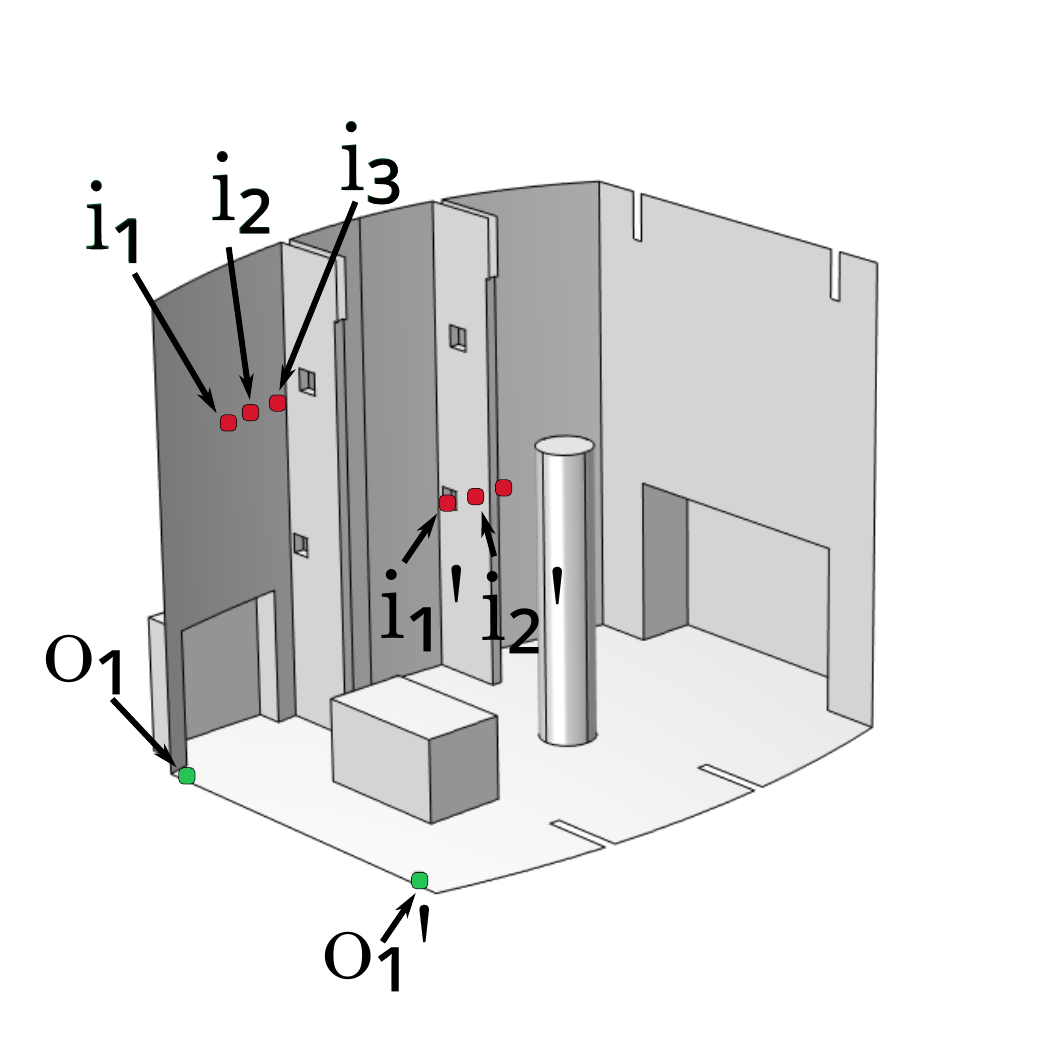}
        \caption{\textit{Configuration 1}}
        \label{fig:src1}
    \end{subfigure}
    \hspace{0.1cm}
    \begin{subfigure}[b]{0.24\linewidth}
        \includegraphics[width = \linewidth]{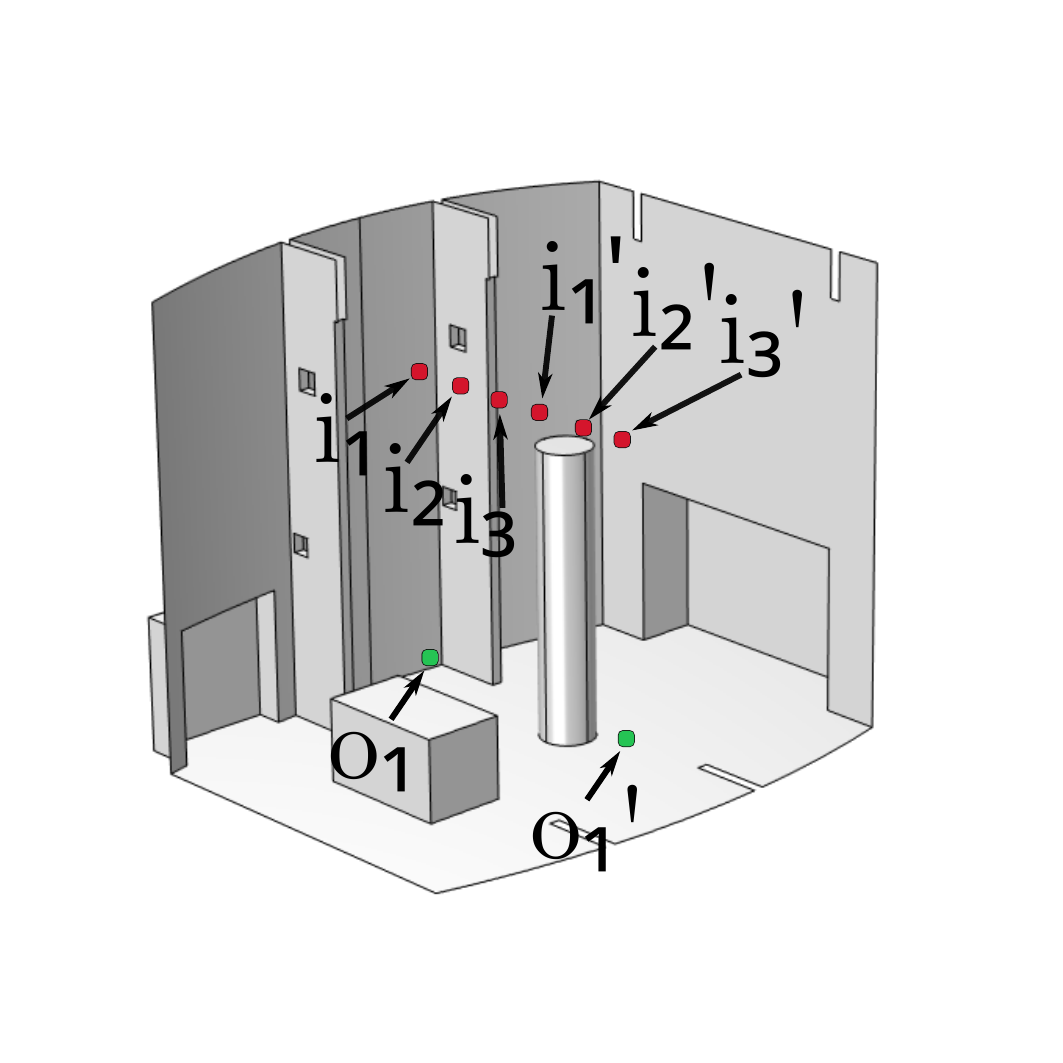}
        \caption{\textit{Configuration 2}}
        \label{fig:src2}
    \end{subfigure}
    \begin{subfigure}[b]{0.49\linewidth}
        \includegraphics[width=\textwidth]{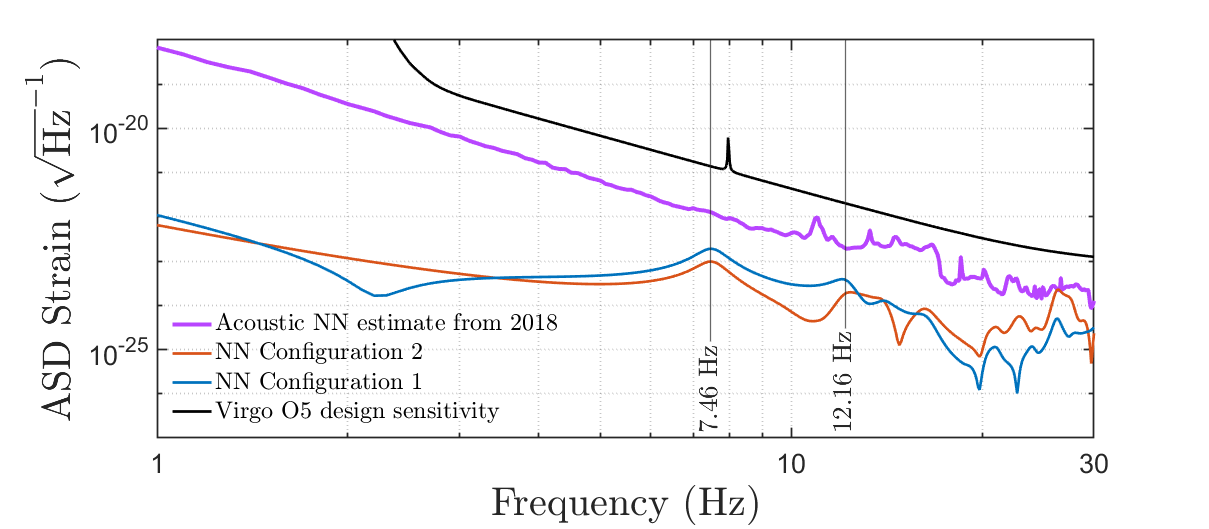}
        \caption{}
        \label{fig:posSRC}
    \end{subfigure}
    \begin{subfigure}[b]{0.49\linewidth}
        \includegraphics[width=\textwidth]{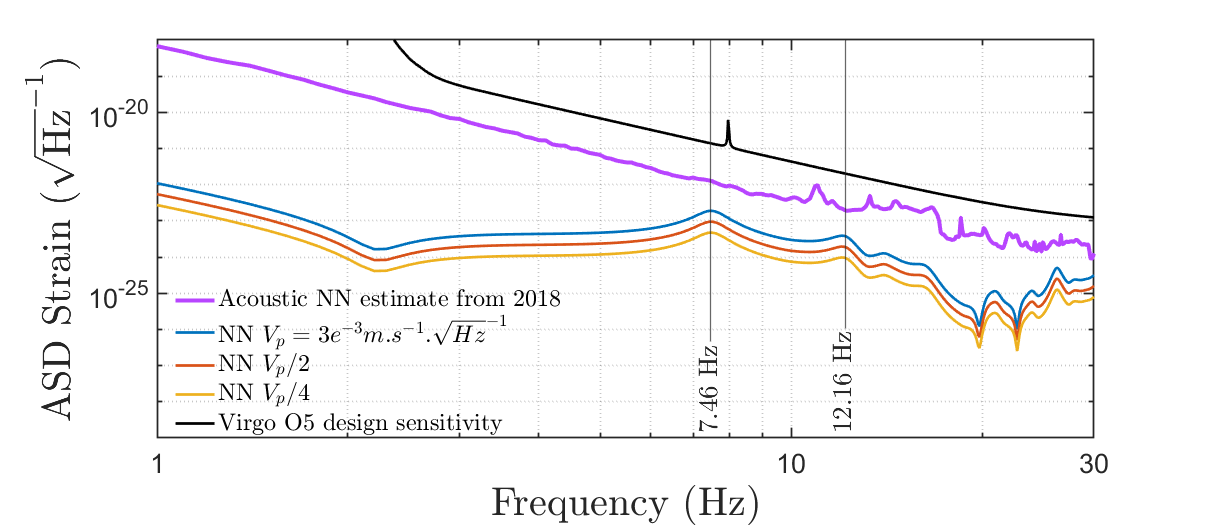}
        \caption{}
        \label{fig:ampSRC}
    \end{subfigure}
    \begin{subfigure}[b]{0.49\linewidth}
        \includegraphics[width=\textwidth]{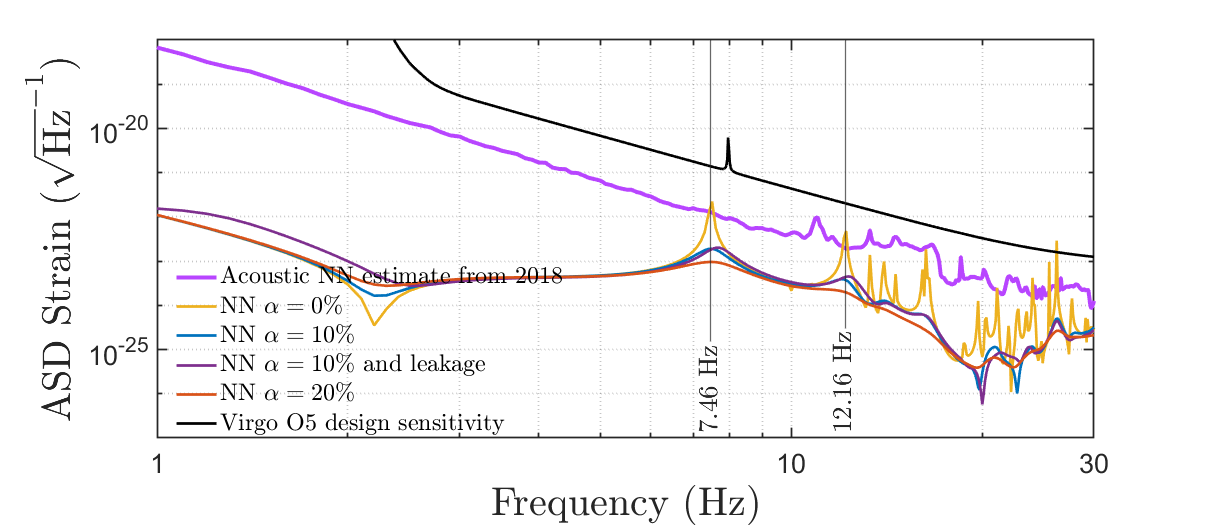}
        \caption{}
        \label{fig:Damp}
    \end{subfigure}
    \caption{\textbf{(a)} Initial location of the acoustic monopoles. \textbf{(b)} Alternative positioning of acoustic sources \textbf{(c)} Strain noise of Newtonian origin for two different location of the acoustic sources. \textbf{(d)} Parametric variation of the amplitude of the source piston. \textbf{(e)} Parametric variation of the modal damping coefficient and study of the effect of leakage. In each plot, the black curve indicates the Virgo O5 design sensitivity and the purple curve is the 2018 Newtonian noise estimate from \cite{Fiorucci_2018}. The 2018 estimate takes into account contributions from 3 buildings of the interferometer and from atmospheric Newtonian noise that are not considered in the model. In this paper only internal acoustic contribution of the 3 buildings is being studied.}
    \label{fig:param}
\end{figure*}

Describing the acoustic source as a set of monopoles makes it easy to perform all kinds of parametric analyzes. In this section, three methods are investigated to reduce the ambient acoustic perturbation and, therefore, to reduce the Newtonian noise it generates. First, the effect of modifying the source location is analyzed, secondly, the effect of the source's amplitude is examined and lastly, the impact of the wall's absorption is observed. Regarding the observation of the absorption coefficient effect, a side study including small leakage of the room is also described. The results are shown in Fig. \ref{fig:param}.\medskip

As the room's geometry cannot be easily modified in practice, its modal behavior cannot be changed. To reduce the acoustic coupling between a source and the room's modes, one can try to optimize the location at which it is positioned \cor{with respect} to the spatial distribution of the mode. The goal is to locate the acoustic sources as close as possible to acoustic pressure nodes. In the case of Virgo, the inlets and outlets are distributed throughout the room as a means to effectively convey air, and their position is not optimized from an acoustic point of view. Even more problematic, the inlets that radiate the most are positioned in the corners of the room, where all of the acoustic modes can couple due to the presence of pressure antinodes. Fig. \ref{fig:src1} and \ref{fig:src2} show the two possible configurations of the sources studied. The assumption is made that the source's acoustic flow does not change with a modification of the duct's geometry. This is justified by the fact that due to the strong absorption coefficient of $\alpha=0.08$, the ducts are non-resonant, as can be seen in the air flow computation. The effect this modification has on the Newtonian noise is displayed in Fig. \ref{fig:posSRC}, in which the total noise and the 2018 estimate of the Newtonian noise from \cite{Fiorucci_2018} shown in Fig. \ref{fig:1} are plotted. The gravity field shows that modes $(1,0,0)$ and $(1,1,0)$ have the most impact, as their main component lies in the x direction. This is recurrent in all the following plots. For now, the position of the sources is arbitrary and aims to position all the sources on the pressure node of the $(1,1,0)$ mode. It is foreseeable to describe an optimization scheme in order to reduce the ambient acoustic and Newtonian noise by finding the adequate positioning of the sources. Another consideration to be taken into account is the working point of the HVAC system. A diminution in the air flow of the heating ventilation and air conditioning system leads to a straight reduction in the acoustic level inside of the room as shown in \cite{Accadia}. This straightforward effect is displayed in Fig. \ref{fig:ampSRC} where 3 different amplitudes are observed, the first curve represents the nominal source amplitude and the two following ones are respectively displaying a source that has been halved and one that has been quartered. The last effect studied is the impact of the absorption coefficient on the behavior of the room. This absorption coefficient could be increased in order to attenuate the acoustic modal components and, consequently, Newtonian noise. The corresponding parametric study is displayed in Fig. \ref{fig:Damp}. The low frequency modes are particularly difficult to fend off and classical methods such as foam and resonators cannot cover such a frequency band out of practicality.

\medskip

In Fig. \ref{fig:param}, the estimated Newtonian noise from \cite{Fiorucci_2018} is higher than the proposed model. Our prediction takes into account the internal contribution coming from three buildings of the interferometer, whereas \cite{Fiorucci_2018} accounts for the internal contribution from the three buildings housing the interferometer and the external contribution measured at the Virgo Central Building (CEB) for all four test masses. According to the 2018 paper, the main contribution to Newtonian noise below 10 Hz comes from the external part (e.g. acoustic waves of the atmosphere). In \cite{Fiorucci_2018}, the estimated external contribution below 10 Hz
is said to be a worst-case scenario as noise generated by wind was limiting the
sensitivity of the microphones used. In addition, the Virgo detector underwent several upgrades after 2018. Among the work performed, the HVAC units of both end buildings (NEB and WEB) were upgraded and the acoustic noise in the terminal building was reduced by up to a factor 4 below 10 Hz as shown in \cite{Tringali_2026}. Measurements performed in the context of the present study were taken after the reduction in acoustic noise.

At 10 Hz, the amplitude of the Newtonian noise reaches $2.9\times10^{-24}\ \mathrm{\sqrt{Hz}^{-1}}$, which is $\approx1.4\times10^{2}$ times smaller than the O5 design sensitivity. The acoustic Newtonian noise is therefore not a limiting factor for Virgo as the relative amplitude between design sensitivity and estimate is at least of a factor 10 over the whole frequency range of interest.%

\section{Conclusion\label{sec:4-conclusion}}

Newtonian noise results from density variations in the mediums surrounding the test masses of the interferometer. Here, we look at such variations generated by the ambient acoustic field of the experimental area in \cor{the end buildings of the detector. In those buildings,} the main source of acoustic noise is the heating ventilation and air conditioning (HVAC) system. The acoustic field is assessed through numerical simulation using finite elements in Comsol Multiphysics software. As Newtonian noise could limit the detector's sensitivity below a few tens of Hertz and because the modal density is high, we limit the range of the computation to the 1-30 Hz frequency band. The wall surfaces are supposed to have an acoustic absorption coefficient, estimated from the measurement of the RT60 reverberation time and Sabine's law. \\

Two models are considered :
\\
The first includes the ducts that carry air and sound from the ventilation system's propeller to the experimental hall. This version gives us the ability to compute the acoustic transfer function of the room with great accuracy, as shown in Fig. \ref{fig:3}. \cor{This model can then be scaled to match the experimental data to extract an equivalent acoustic source to represent the HVAC system.} This model requires a high computation time and imposes strong constraints on performing parametric studies. This leads to the consideration of the second, simpler model. We use the low frequency approximation to describe the radiating acoustic elements inside the experimental hall as a set of monopoles. Their amplitudes (\textit{i.e.} Acoustic Flow) are deduced from the computed transfer function using the complete model and experimental data. This reduced complexity gives us the possibility to perform a set of parametric studies regarding the location of sources, \cor{their relative phases and amplitudes and the acoustic absorption of the room. The eigenfrequencies and their associated eigenshapes are then computed using the simplified geometry. Using a gauss quadrature algorithm, the functions associated to the potential and gravity field are then computed. The acoustic and gravitational field generated by the set of monopoles can then be reconstructed using a modal basis projection}. The model shows us that the inlets and outlets that radiate the most are those located close to the propeller. Their positioning in the corners of the room eases the coupling between the HVAC and the acoustic modes of the experimental area. Some acoustic modes of the room have a stronger impact on the Newtonian noise level of the interferometer due to their spatial distribution. The first and fourth modes of the room seem to generate a perturbation aligned with the arm of the detector, implying strong coupling.\medskip

This article presents a methodology for assessing the distribution and amplitude of the gravity field inside a room when it is generated by an acoustic perturbation. The novelty of this approach lies in the fact that the acoustic pressure field is described precisely using modal superposition and the finite element method. This gives us the ability to optimize the shape and absorption of the room, as well as the location of the acoustic sources and sensitive apparatus inside. The estimated Newtonian noise level can help define the maximum sound power criteria for the design of an improved HVAC system. This would contribute to the design of future detectors such as the European Einstein Telescope \cite{fenyvesiMitigationEffectChanges2024} project. 

Compared to the estimate given in \cite{Fiorucci_2018}, the acoustic Newtonian noise
calculated in this paper remains one order of magnitude lower (strain at $10 \mathrm{Hz}$
equal to $2.9\times40^{-24}$ for configuration 1 in Fig \ref{fig:posSRC}, and $4.3\times10^{-23}$ for the estimate in \cite{Fiorucci_2018}). In addition to the different modeling approach, there are two other possible reasons for this difference: first, the estimate reported in \cite{Fiorucci_2018} was obtained before the improvements made to the HVAC in 2018. Consequently, the buildings were noisier than they are in the measurements used in the present paper. Second, the estimate in \cite{Fiorucci_2018} accounts for Newtonian noise induced by both internal acoustic phenomena and external atmospheric phenomena. As noted in \cite{Cafaro_2009} and \cite{Creighton_2008}, atmospheric phenomena can also be a source of Newtonian noise. For example, air convection around buildings, convected temperature pockets, or acoustic waves induced by turbulence contribute to the generation of Newtonian noise; their contributions are not accounted for in this study.

As Newtonian noise is impossible to completely mitigate, active methods to cancel it are already being developed for Virgo \cite{Koley_2024,Harms_LowerLimitnewtoniannoise_2022a,Singha_CharacterizationSeismicField_2021,Bader_SeismicnewtonianNoise_2021}. When it comes to acoustics, one must keep in mind that Newtonian noise is only one of the expected noise coupling mechanisms, among others, such as scattered light, beam jitter, or optical fiber noise.

\section{Acknowledgment}

\cor{The authors wish to give their gratitude to the European Gravitational Observatory (EGO) for providing access to the microphones data and for their support in this work. Tomasz Bulik was supported by the European Union's Horizon Europe research and innovation programme under the grant agreement No. 101079696 (HORIZON-INFRA-2021-DEV-02 (ET-PP)) and by the Polish Ministry of Science and Higher Education project (W55/HE/2022). The authors also wish to thank Le Mans University for the funding of this work.} 

\appendix

\section*{Appendix}

\section{Acoustic measurement setup in the WEB \label{sec-5A}}

\cor{This appendix describes the experimental setup used to validate the proposed acoustic model. It consists of an array of 22 microphones installed inside the WEB. The location of each microphone used is indicated in Fig. \ref{fig:3b}: The blue dots represent the sensors, that are shown in this paper. Microphones of the array are named by numbers ranging from 1 to 22. The microphones are custom made. As reported in \cite{fenyvesiMitigationEffectChanges2024}, the frequency range of operation is 0.1 Hz to 120 Hz with a sensitivity of 2.5 mV/Pa. The sensors are connected to a standalone acquisition system (sample frequency 500 Hz) and then synchronized with the Virgo data-stream.}

Two commercial microphones are also used (B\&K 4193 type, frequency range 0.07 Hz - 20 kHz, sensitivity 12.5 mV/Pa, dynamic range 19 - 162 dB). They are connected to the Virgo data acquisition system (DAQ \cite{Letendre_2017}). One of them is fixed and located at 6 meters from the ground at location labeled MIC in Fig. \ref{fig:3b}.

\begin{figure}[H]
\centering
\begin{subfigure}[b]{\linewidth}
    \includegraphics[width=\linewidth]{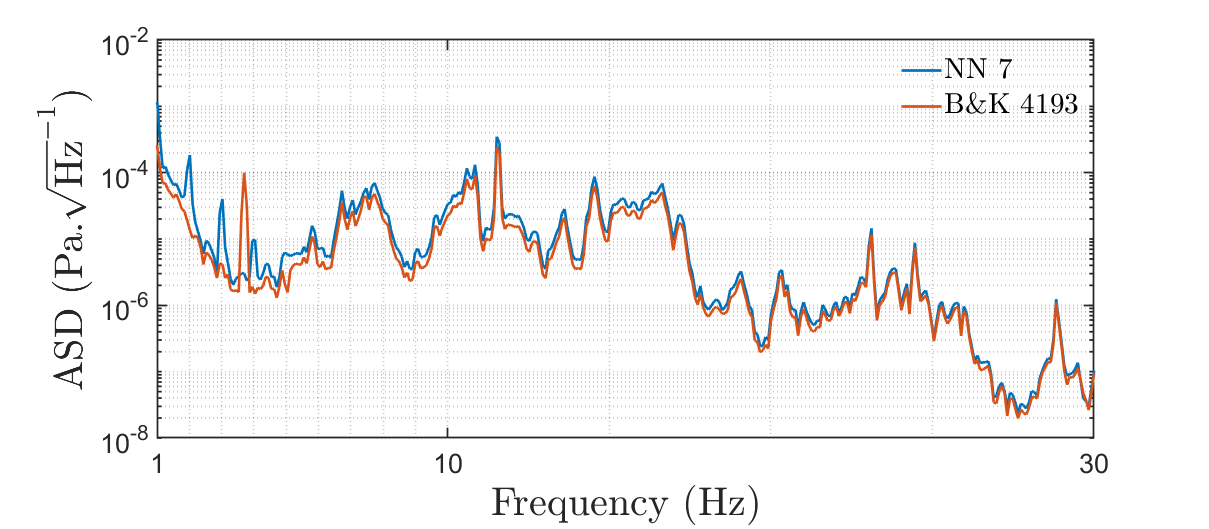}
    \caption{}
    \label{fig:ASD}
\end{subfigure}
\begin{subfigure}[b]{\linewidth}
    \includegraphics[width=\linewidth]{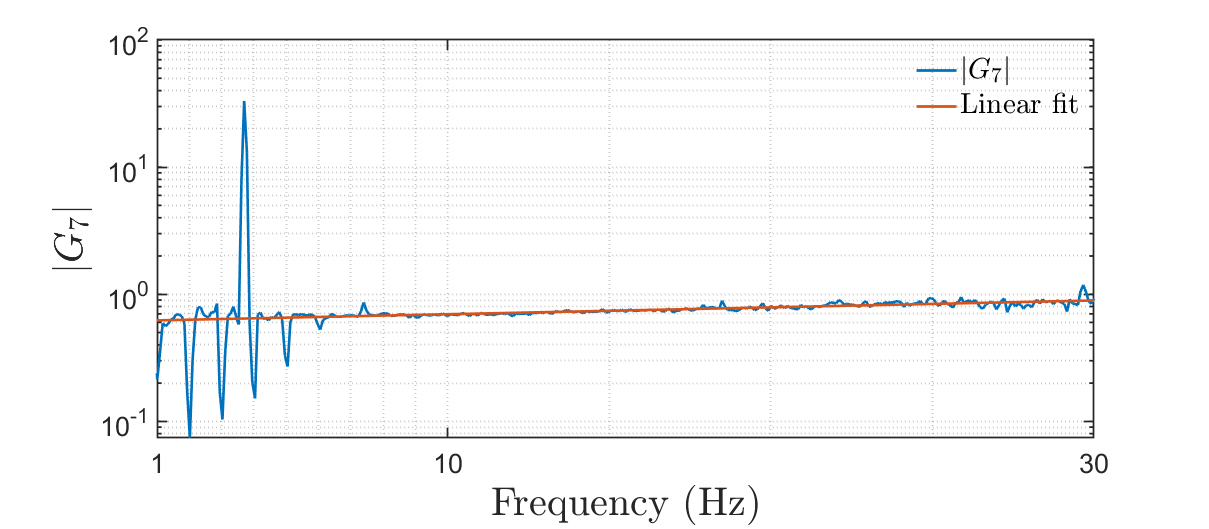}
    \caption{}
    \label{fig:H}
\end{subfigure}
\begin{subfigure}[b]{\linewidth}
    \includegraphics[width=\linewidth]{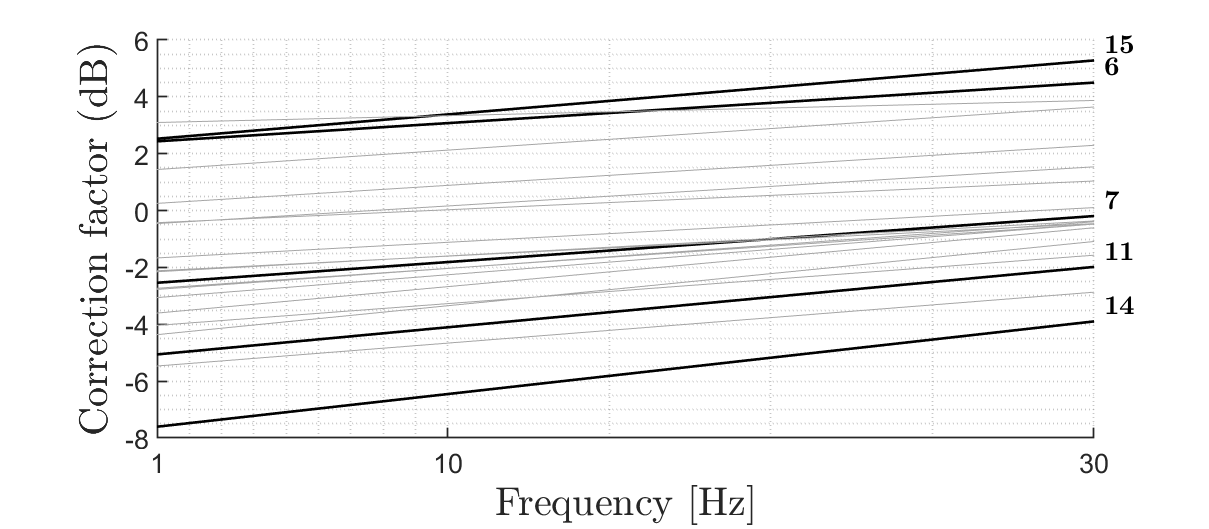}
    \caption{}
    \label{fig:calibrationcurves}
\end{subfigure}
\caption{\textbf{(a)} ASD of microphone NN 7 and the reference B\&K microphone. \textbf{(b)} Modulus of transfer function $|G_7|$ and linear fit of this function. \textbf{(c)} Correction factor estimated for the whole sensor array. Some curves are highlighted, the name of the microphone for which they are estimated is shown to the right. RefdB = 1.}
\label{fig:calibration}
\end{figure}

To ensure the accuracy of the data, the entire microphone array is calibrated using the second B\&K 4193 microphone, which serves as a mobile reference. This roving reference is placed a few centimeters away from each microphone, in order to compute the transfer function $G_k=p_k/p_{ref}$ ($p_k$ and $p_{ref}$ being the measured pressure by microphone $k$ and by the reference). The ASD for microphone NN 7 and of the reference microphone is shown in Fig. \ref{fig:ASD}. The modulus of the transfer function $G_7$ of one microphone taken as an example (microphone 7) and a linear fit are shown in Fig. \ref{fig:H}. The linear fit of $|G_k|$ serves as a correction factor applied to microphone $k$ measurements to compensate for differences in sensitivity among the sensors (see Fig. \ref{fig:calibrationcurves}).

\section{Variability of the acoustic pressure spectra over a six month period \label{sec-5B}}

\begin{figure}
    \includegraphics[width=\linewidth]{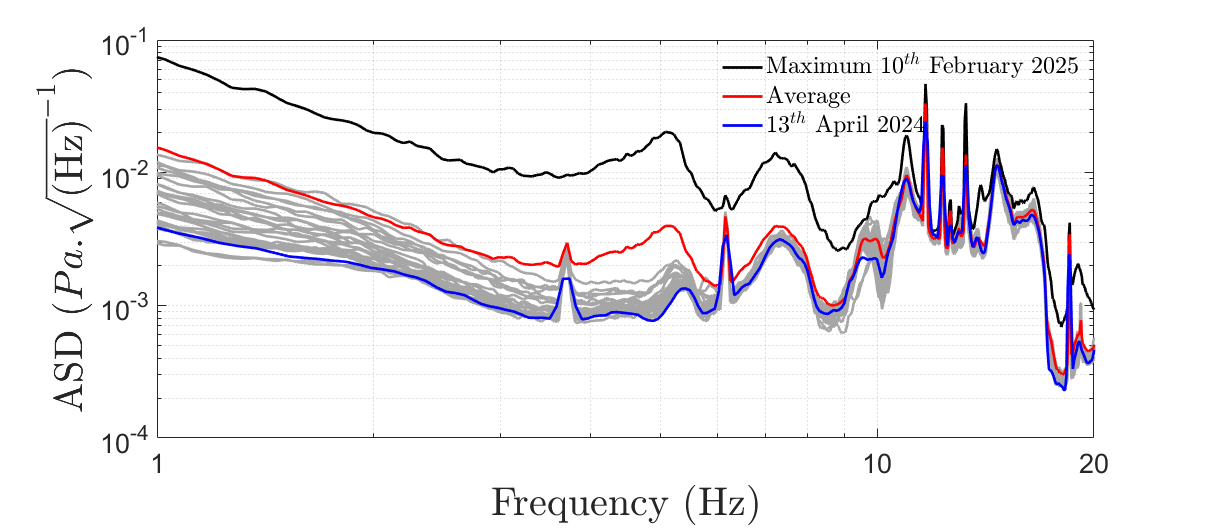}
    \caption{\corsec{Grey curves represent acoustic pressure ASD measured each Monday} for one hour, over a six-month period from September 2, 2024, to March 24, 2025. \corsec{Data are taken between 00:00:00 and 01:00:00 UTC. 10 seconds long Hann windows with 50\% overlap are used to compute the ASD using Welch method. Measurements are performed at microphone MIC location in the WEB. Only the maximum curve is highlighted in black as it represents an anomalously high level of acoustic noise. Red curve is the average of the weekly ASD over the 6 months period. Blue curve represents the ASD shown in Fig. \ref{fig:5} that is used to estimate the equivalent acoustic source $V$, it is not taken in computation of the average represented by the red curve as it is part of another dataset.}}
    \label{fig:ENVtime}
\end{figure}

\cor{The sound pressure spectrum used to estimate the piston speed was measured on April 13, 2024. This choice is arbitrary, but it is representative of a quiet, normal environment with no unusual events. The variability of this spectrum over a six-month period is illustrated in Fig. \ref{fig:ENVtime}. The MIC microphone shown in Fig. \ref{fig:3b} is used for the monitoring. The data are sampled at 50 Hz. Below 5 Hz, the sound pressure level can vary by a factor of 5. Variability above 5 Hz remains low. It should be noted that the HVAC system settings (fan speed) have not changed: this can be seen in the tonal components at 11 Hz, 12.3 Hz, and 13.3 Hz, which remain constant throughout the entire period. We observe that the slight peak at 5.4 Hz varies in both frequency and amplitude over time. We hypothesized that it corresponded to a wind-induced vibration, in one of the building’s walls.}

\bibliographystyle{unsrt}
\bibliography{biblio}

@article{Virgo_2015,
	title = {Advanced {Virgo}: a second-generation interferometric gravitational wave detector},
	volume = {32},
	issn = {0264-9381, 1361-6382},
	doi = {10.1088/0264-9381/32/2/024001},
	language = {en},
	number = {2},
	urldate = {2024-12-04},
	journal = {Classical and Quantum Gravity},
	author = "F. Acernese et al.",
	year = {2015},
}

@article{Ligo_2015,
	title = {Advanced {LIGO}},
	volume = {32},
	issn = {0264-9381, 1361-6382},
	doi = {10.1088/0264-9381/32/7/074001},
	number = {7},
	journal = {Classical and Quantum Gravity},
	author = "The LIGO Scientific Collaboration",
	month = apr,
	year = {2015},
}

@article{Kagra_2020,
  title = {Overview of {{KAGRA}}: {{Detector}} Design and Construction History},
  author = "T. Akutsu et al.",
  year = 2020,
  journal = {Progress of Theoretical and Experimental Physics},
  volume = {2021},
  number = {5},
  pages = {05A101},
  issn = {2050-3911},
  doi = {10.1093/ptep/ptaa125},

}

@techreport{Accadia,
    author    = "T. Accadia et al.",
    type      = {Tech. rep.},
    institution    = "Virgo Technical documentation system",
    title     = "{Advanced Virgo Technical Design Report}",
    number    = "VIR-0128A-12",
    year      = "2012"
}

@article{LIGOScientific:2025slb,
  title = {{{GWTC-4}}.0: {{An Introduction}} to {{Version}} 4.0 of the {{Gravitational-Wave Transient Catalog}}},
  shorttitle = {{{GWTC-4}}.0},
  author = {{The LIGO Scientific Collaboration, the Virgo Collaboration, and the KAGRA Collaboration}},
  year = 2025,
  journal = {The Astrophysical Journal Letters},
  volume = {995},
  number = {1},
  pages = {L18},
  issn = {2041-8205, 2041-8213},
  doi = {10.3847/2041-8213/ae0c06},
}

@article{Fiorucci_2018,
	title = {Impact of infrasound atmospheric noise on gravity detectors used for astrophysical and geophysical applications},
	volume = {97},
	issn = {2470-0010, 2470-0029},
	doi = {10.1103/PhysRevD.97.062003},
	language = {en},
	number = {6},
	journal = {Physical Review D},
	author = "D. Fiorucci et al.",
	year = {2018},
}

@book{Maggiore_2007,
    author = "M. Maggiore",
    title = {Gravitational Waves: Volume 1: Theory and Experiments},
    publisher = {Oxford University Press},
    year = {2007},
    isbn = {9780198570745},
    doi = {10.1093/acprof:oso/9780198570745.001.0001},
}

@book{Kuttruff_2009,
	edition = {$5^{th}$},
	title = {Room acoustics},
	isbn = {978-0-415-48021-5 978-0-203-87637-4},
	language = {en},
	publisher = {CRC Press},
	author = "H. Kuttruff",
    address = "London, UK",
	year = {2009},
}

@article{Jiang,
    title = {A literature review on the squirrel-cage fans using in HVAC equipment: Powerful, efficient, and quiet operation},
    journal = {Journal of Building Engineering},
    volume = {73},
    pages = {106691},
    year = {2023},
    issn = {2352-7102},
    doi = {https://doi.org/10.1016/j.jobe.2023.106691},
    author    = "B. Jiang et al.",
}

@article{neiseNoiseReductionCentrifugal1976,
  title = {Noise Reduction in Centrifugal Fans: {{A}} Literature Survey},
  author = {Neise, W.},
  year = {1976},
  journal = {Journal of Sound and Vibration},
  volume = {45},
  number = {3},
  pages = {375--403},
  doi = {10.1016/0022-460X(76)90394-1},
  langid = {english},
}

@book{Bruneau_Scelo_2006,
    author    = "M. Bruneau and T. Scelo",
    title     = "Fundamentals of acoustics",
    year      = "2006",
    publisher = "ISTE Ltd",
    address   = "London, UK",
    ISBN = "9781905209255"
}

@book{Fahy_Gardonio_2007,
  title = {Sound and Structural Vibration: Radiation, Transmission and Response},
  author = "F. Fahy and J. Frank and P. Gardonio",
  year = 2007,
  edition = {2nd ed},
  publisher = {Elsevier/Academic},
  address = {Amsterdam London},
  isbn = {978-0-12-373633-8},
  langid = {english},
  lccn = {620.2},
}

@article{Maria_2020,
  title = {Seismic Array Measurements at {{Virgo}}’s West End Building for the Configuration of a {{Newtonian-noise}} Cancellation System},
  author = "M. Tringali et al.",
  year = "2020",
  journal = {Classical and Quantum Gravity},
  volume = {37},
  number = {2},
  issn = {0264-9381, 1361-6382},
  doi = {10.1088/1361-6382/ab5c43},
}

@article{Trozzo_2022,
  title = {Seismic and {{Newtonian Noise}} in the {{GW Detectors}}},
  author = "L. Trozzo and F. Badaracco",
  year = "2022",
  journal= {Galaxies},
  volume = {10},
  number = {1},
  pages = {20},
  issn = {2075-4434},
  doi = {10.3390/galaxies10010020},
  langid = {english},
}

@article{Koley_2024,
  title = {Design and Implementation of a Seismic {{Newtonian}} Noise Cancellation System for the {{Virgo}} Gravitational-Wave Detector},
  author = "S. Koley et al.",
  year = "2024",
  journal = {The European Physical Journal Plus},
  volume = {139},
  number = {1},
  pages = {48},
  issn = {2190-5444},
  doi = {10.1140/epjp/s13360-023-04834-0},
  langid = {english},
}

@techreport{Falxa_2018,
	title = {Acoustic characterization of {Advanced} {Virgo} buildings},
	language = {en},
	author = "M. Falxa et al.",
    type      = {Tech. rep.},
    institution    = "Virgo Technical documentation system",
    number    = "VIR-0673A-18",
    year = {2018}
}

@phdthesis{Bader_SeismicNewtonianNoise_2021,
    author = "M. K. M. Bader",
    title = "{Seismic and Newtonian noise modeling for Advanced Virgo and Einstein Telescope}",
    school = "Vrije U., Amsterdam, Vrije U., Amsterdam",
    year = "2021"
}

@techreport{Dandrea_TechnicalReportPlanning_2021,
    title = {Technical {Report} and {Planning} - {Activities} for {Infrastructure} {HVAC} -},
    author = "M. D’Andrea et al",
    type      = {Tech. rep.},
    institution    = "Virgo Technical documentation system",
    number    = "VIR-0181C-21",
    year = {2021}
}

@article{Singha_CharacterizationSeismicField_2021,
	title = {Characterization of the seismic field at {Virgo} and improved estimates of {Newtonian}-noise suppression by recesses},
	volume = {38},
	issn = {0264-9381, 1361-6382},
	doi = {10.1088/1361-6382/ac348a},
	number = {24},
	journal = {Classical and Quantum Gravity},
	author = "A. Singha et al.",
	year = {2021},
}

@article{Harms_LowerLimitNewtoniannoise_2022a,
    title = {A lower limit for {Newtonian}-noise models of the {Einstein} {Telescope}},
    author = {J. Harms et al.},
    year = 2022,
    journal = {The European Physical Journal Plus},
    volume = {137},
    number = {6},
    pages = {687},
    issn = {2190-5444},
    doi = {10.1140/epjp/s13360-022-02851-z},
    langid = {english},
}

@article{Harms_TerrestrialGravityFluctuations_2019,
	title = {Terrestrial gravity fluctuations},
	volume = {22},
	issn = {2367-3613, 1433-8351},
	doi = {10.1007/s41114-019-0022-2},
	language = {en},
	number = {1},
	journal = {Living Reviews in Relativity},
	author = {J. Harms},
	month = dec,
	year = {2019},
}

@misc{Cafaro_2009,
	title = {Analytical {Estimate} of {Atmospheric} {Newtonian} {Noise} {Generated} by {Acoustic} and {Turbulent} {Phenomena} in {Laser}-{Interferometric} {Gravitational} {Waves} {Detectors}},
	doi = {10.48550/arXiv.0906.4844},
	language = {en},
	publisher = {arXiv},
	author = {C. Cafaro and S. A. Ali},
    journal = "arXiv:0906.4844v2",
	year = {2009},
}

@article{Saulson_1984,
	title = {Terrestrial gravitational noise on a gravitational wave antenna},
	volume = {30},
	issn = {0556-2821},
	doi = {10.1103/PhysRevD.30.732},
	language = {en},
	number = {4},
	journal = {Physical Review D},
	author = {P. R. Saulson},
	year = {1984},
}

@article{Creighton_2008,
	title = {Tumbleweeds and airborne gravitational noise sources for {LIGO}},
	volume = {25},
	issn = {0264-9381, 1361-6382},
	doi = {10.1088/0264-9381/25/12/125011},
	language = {en},
	number = {12},
	journal = {Classical and Quantum Gravity},
	author = {T. Creighton},
	year = {2008},
	pages = {125011},
}

@article{Fiori_2020,
  title = {The {{Hunt}} for {{Environmental Noise}} in {{Virgo}} during the {{Third Observing Run}}},
  author = "I. Fiori et al.",
  year = 2020,
  journal = {Galaxies},
  volume = {8},
  number = {4},
  issn = {2075-4434},
  doi = {10.3390/galaxies8040082},
  langid = {english},
}

@article{Basti_2023,
title = {The seismic isolation system of Advanced Virgo Plus, Phase II},
journal = {Nuclear Instruments and Methods in Physics Research Section A: Accelerators, Spectrometers, Detectors and Associated Equipment},
volume = {1048},
pages = {168021},
year = {2023},
issn = {0168-9002},
doi = {https://doi.org/10.1016/j.nima.2023.168021},
author = "A. Basti et al."}

@misc{gwinc,
    title = {Gravitational Wave Interferometer Noise (and inspiral range) Calculators (V0.6.2) URL : https://git.ligo.org/gwinc},
    year = {2024}
}

@article{fenyvesiMitigationEffectChanges2024,
  title = "{Mitigation of the effect of changes of atmospheric pressure on gravity detectors: preliminary results obtained with microphones at the Sos Enattos mine}",
  author = "E. Fenyvesi et al.",
  doi = "10.22323/1.441.0104",
  journal = "PoS",
  year = 2024,
  volume = "TAUP2023",
  pages = "104"
}

@techreport{Letendre_2017,
    author = "N. Letendre et al.",
    title = "{ADC}7674 user Manual {V}2",
    type = {Tech. rep.},
    institution = "Virgo Technical documentation system",
    number = "VIR-0937A-17",
    year = {2017}
}

@article{Tringali_2026,
  title = {Reducing the {{Virgo}} Site Infrastructure Noise in Preparation of the {{O4}} Observing Run},
  author = "I. Fiori et al.",
  year = {2026},
  eprint = {2603.28960},
  eprinttype = {arXiv},
  doi = {10.48550/arXiv.2603.28960},
  langid = {english},
}

\end{document}